\documentclass{emulateapj}

\usepackage{amsmath}
\usepackage{amssymb}
\usepackage{graphicx}
\usepackage{color}
\newcommand{\lSect}[1]{{\label{sec:#1}}}

\def\gtaprx {\lower .1ex\hbox{\rlap{\raise .6ex\hbox{\hskip .3ex
	{\ifmmode{\scriptscriptstyle >}\else
		{$\scriptscriptstyle >$}\fi}}}
	\kern -.4ex{\ifmmode{\scriptscriptstyle \sim}\else
		{$\scriptscriptstyle\sim$}\fi}}}
\def\ltaprx {\lower .1ex\hbox{\rlap{\raise .6ex\hbox{\hskip .3ex
	{\ifmmode{\scriptscriptstyle <}\else
		{$\scriptscriptstyle <$}\fi}}}
	\kern -.4ex{\ifmmode{\scriptscriptstyle \sim}\else
		{$\scriptscriptstyle\sim$}\fi}}}

\newcommand{\cutt}[1]{\textcolor{blue}{}}

\newcommand{\code}[1]{{\tt{#1}}}
\newcommand{\Msun}{{\ensuremath{\mathrm{M}_{\odot} }}}
\newcommand{\Zsun}{\ensuremath{Z_\odot}}

\newcommand{\Ni}{{\ensuremath{^{56}\mathrm{Ni}}}}

\newcommand{\Co}{{\ensuremath{^{56}\mathrm{Co}}}}

\newcommand{\N}{{\ensuremath{^{14} \mathrm{N}}} }

\newcommand{\Sectff}[1]{{\ref{sec:#1}}}
\newcommand{\Sect}[1]{{\S~\Sectff{#1}}}

\begin{document}

\title{Primordial Core-Collapse Supernovae and the Chemical Abundances
  of Metal-Poor Stars}

  \author{C.C.  Joggerst,\altaffilmark{1,2}
    A. Almgren,\altaffilmark{3} J. Bell,\altaffilmark{3}, Alexander
    Heger,\altaffilmark{4} Daniel Whalen\altaffilmark{2} and
    S. E. Woosley\altaffilmark{1}}

\altaffiltext{1}{Department of Astronomy and Astrophysics,
University of California at Santa Cruz, Santa Cruz, CA 
95060; cchurch@ucolick.org}

\altaffiltext{2}{Theoretical Astrophysics (T-2), Los Alamos National
Laboratory, Los Alamos, NM 87545}

\altaffiltext{3}{Computational Research Division, Lawrence Berkeley 
National Lab, Berkeley, CA 94720}

\altaffiltext{4}{School of Physics and Astronomy, University of 
Minnesota, Twin Cities, Minneapolis, MN 55455}

\begin{abstract}
The inclusion of rotationally-induced mixing in stellar evolution can
alter the structure and composition of presupernova stars.  We survey
the effects of progenitor rotation on nucleosynthetic yields in
Population III and II supernovae using the new adaptive mesh
refinement (AMR) code \code{CASTRO}.  We examine piston-driven
spherical explosions in 15, 25 and 40 \Msun \ stars at $Z = 0$ and
$10^{-4}$ \Zsun \ with three explosion energies and two rotation
rates.  Rotation in the $Z = 0$ models resulted in primary nitrogen
production and a stronger hydrogen burning shell which led all models
to die as red supergiants (in contrast to the blue supergiant
progenitors made without rotation).  On the other hand, the
$Z=10^{-4}$ \Zsun \ models that included rotation ended their lives as
compact blue stars. Because of their extended structure, the
hydrodynamics favors more mixing and less fallback in the metal free
stars than the $Z = 10^{-4}$ models.  As expected, higher energy
explosions produce more enrichment and less fallback than do lower
energy explosions, and at constant explosion energy, less massive
stars produce more enrichment and leave behind smaller remnants than
do more massive stars.  We compare our nucleosynthetic yields to the
chemical abundances in the three most iron-poor stars yet found and
reproduce the abundance pattern of one, HE 0557-4840, with a zero
metallicity 15 \Msun, $2.4 \times 10^{51}$ erg supernova.  A Salpeter
IMF averaged integration of our yields for $Z=0$ models with explosion
energies of $2.4 \times 10^{51}$ ergs or less is in good agreement
with the abundances observed in larger samples of extremely metal-poor
stars, provided 15 \Msun \ stars are included.  Since the abundance
patterns of extremely metal-poor stars likely arise from a
representative sample of progenitors, our yields suggest that 15-40
\Msun \ core-collapse supernovae with moderate explosion energies
contributed the bulk of the metals to the early universe.

\end{abstract}

\keywords{Supernovae, nucleosynthesis, first stars}

\maketitle

\section{Introduction}
\lSect{introduction}


Early chemical enrichment of the universe began with the deaths of the
first stars at $z \sim$ 20 - 30
\citep{mbh03,ky05,greif07,kjb08,wet08}.  The nature of the first
stars, and hence the primordial initial mass function (IMF), is yet to
be observationally constrained, but numerical models that proceed from
well-posed cosmological initial conditions suggest that they are very
massive, from 25 - 500 \Msun \citep{bcl99,
bcl02,nu01,abn00,abn02,on07,wa07}.  The non-rotating stellar evolution
models of \citet{Heger&Woosley2002} predict that 15 - 40 \Msun \ Population III
(Pop III) stars die in conventional core-collapse supernovae, while
140 - 260 \Msun \ stars explode as much more energetic
pair-instability supernovae.  For this reason, 15 - 40 \Msun \ Pop III
stars are considered to be "low-mass" primordial stars and 140 - 260
\Msun \ stars are termed "very massive" stars.  Although direct
observation of the first stars is not yet possible, clues to their
nature may be extracted from their nucleosynthetic imprint on
second-generation stars.  The gas enriched by Population III stars is
thought to be able to form low-mass stars that may still exist,
retaining the chemical imprint of their progenitors.  Fossils of this
second generation of stars are now being sought in surveys of
extremely metal-poor (EMP) and hyper metal-poor (HMP) stars in the
Galactic halo \citep{Beers&Christlieb2005, Frebel2005}. HMP stars,
with [Fe/H] $<-4$, are thought to be enriched by only one or a few
SNe.  EMP stars, with $-4 < $[Fe/H]$<-3$, are more plentiful and show
a smaller scatter in their abundance ratios, implying that they formed
from gas that had been enriched by a representative sample of SNe.

Many attempts have been made to reproduce the abundance patterns in
EMP and HMP stars by modeling the evolution and explosion of massive
stars, and comparing the yields to observations \citep{Nomoto2006,
Umeda&Nomoto2002, Umeda&Nomoto2003,Umeda&Nomoto2005, Joggerst2009},
and in doing so, gain a better understanding of the nature of the
first stars. Some studies have focused on the potential role of very
massive stars that die as pair-instability supernovae; others suggest
that very energetic hypernova explosions may be responsible for these
abundance patterns.  Although constraints have been placed on the
energy and origin of the explosion within the star and theoretical
progress has been made \citep{Burrows2006, Murphy2008, Marek2009}, the
explosion mechanism that operates in this first generation of stars is
not well understood.  Furthermore, core-collapse SNe leave compact
remnants onto which the outer layers of the star may fall, preventing
their escape and enrichment of the next generation of stars.  The
degree to which the layers of the star are mixed by the
Rayleigh-Taylor (RT) instability prior to shock breakout from the
surface of the star, together with the amount of material that falls
onto the central remnant, governs the final yield of the SN.

The large one-dimensional explosion surveys performed to date
\citep{Iwamoto2005, Tominaga2007, Heger2008} are forced to approximate
these inherently multidimensional processes by artificially mixing the
layers of the star and selecting a ``mass cut'' that denotes the line
between the inner material that falls back onto the central compact
object and the outer material that escapes.  Two-dimensional models
have been constructed to obtain more accurate chemical yields for
core-collapse SNe and to constrain one dimensional parameterizations
of these explosions.  \citet{ Tominaga2009} investigated one explosion
mechanism, jet-induced explosions, with two-dimensional simulations of
a 40 $\Msun$ primordial star and found good agreement using
angle-delimited yields to the abundance pattern in one HMP star, but
was not as sucessful in reproducing the abunace patterns in EMP stars.

Previous studies of mixing in core-collapse supernovae were largely
focused on replicating observations of SN1987A.  The most sucessful of
these were the studies of \citet{Kifonidis2003} and
\citet{Kifonidis2006}, which modeled the explosion in more detail and
from earlier times than previous studies.

The first paper in this series \citep{Joggerst2009} modeled spherical
explosions of non-rotating 15 \Msun \ and 25 \Msun \ Population III
stars with the \code{FLASH} code.  These stars ended their lives as
compact blue supergiants, and experienced far less mixing and more
fallback than the larger-radius solar metallicity red supergiants also
modeled in \citet{Joggerst2009}.  The non-rotating Population III
model yields achieved only partial agreement with the abundances in
the most metal-poor stars ever detected.  In particular, they did not
reproduce the amount of nitrogen observed in HMP stars because it was
not present in the initial pre-supernova progenitor models in
sufficient quantities. The zero-metallicity explosions in this suite
also manifested a high degree of fallback and a low degree of mixing,
less than what had previously been assumed in the one-dimensional
models.  These two factors resulted in even greater [C+O/Fe] ratios
than those in the HMP stars, and failed to match the abundance
patterns observed in the larger sample of EMP stars.

Replicating abundances, particularly of N, in EMP and HMP stars may
require rotationally-induced mixing prior to the destruction of the
star. Recent simulations of Population III stellar evolution up to the
point of explosion by \citet{Ekstroem2008} have demonstrated that
rotationally-induced mixing boosts burning in the H-shell after the
star has left the main sequence, increasing \N production by as much
as a factor of $10^6$ and expanding the outer envelope of the star.
The larger outer envelope in rotating models will impact
post-explosion dynamics in the star, allowing mixing to
occur on longer timescales and perhaps producing more enrichment than
in non-rotating models.

In this, the second of a series of papers on elemental yields and
mixing in Population III supernovae, we examine rotating models in
two-dimensional simulations to determine if they are responsible for
the elemental patterns discovered in the EMP/HMP star surveys. These
studies may also be used to constrain estimations of mixing and
fallback for large ensembles of one dimensional supernova models like
those of \citet{Heger2008}.  While these rotating progenitor models do
not take rotationally-induced mass loss into account, and our
explosions are spherical and not asymmetric (as may well be the case)
these effects may have less impact on the stellar yield than the
radius of the star at the time of its death.  Our rotating models die
as significantly larger, redder stars than the non-rotating models of
\citet{Joggerst2009}, and we anticipate that this will lead to
significantly enhanced mixing and reduced fallback, and thus higher
yields, as compared to non-rotating models.

This survey is more complete in other respects.  Our models capture
mixing and fallback for a wider variety of progenitor masses,
explosion energies, rotation rates, and metallicities than the
previous survey.  We include odd-numbered elements that were not
included in our first survey, allowing for more detailed comparisons
with EMP and HMP abundances. The greater number of elements also lays
the groundwork for future light curve and spectroscopic studies of
these supernovae.  Our current study employs a new code,
\code{CASTRO}, which can better handle the larger number of elements
and will be more easily adapted to future studies than \code{FLASH}.

In \Sect{codes} we describe the new multi-dimensional Eulerian 
hydrodynamics code \code{CASTRO} used to model the explosions.  
We describe the one-dimensional \code{KEPLER} models on which 
our simulations are based and outline our parameter survey in 
\Sect{models} and analyze the chemical yields of our simulation 
campaign in \Sect{results}.  These yields are discussed and 
compared with chemical abundance measurements in HMP and EMP 
stars in \Sect{discussion}.  In \Sect{conclusions} we conclude.

\section{Numerical Algorithms}
\lSect{codes}

The explosion models in this study were implemented in two stages.
First, one-dimensional supernova profiles for the progenitors were
computed in the \code{KEPLER} code to capture all explosive
nucleosynthetic burning.  These profiles were then mapped onto a
two-dimensional R-Z (axisymmetric) grid in the new \code{CASTRO}
hydrodynamics code and evolved out to radii where Rayleigh-Taylor (RT)
mixing ceased and the star was expanding essentially homologously.  We
computed thirty-six such models, covering 3 masses, 3 explosion
energies, 2 metallicities, and 2 rotation rates.

\subsection{\code{CASTRO}}
\lSect{CASTRO}

\code{CASTRO} \citep{Almgren2010} is a multi-dimensional Eulerian AMR
hydrodynamics code.  Time integration of the hydrodynamics equations
is based on a higher-order, unsplit Godunov scheme.  \code{CASTRO} can
perform calculations in one dimensional radial, two dimensional
cylindrical, or three dimensional Cartesian coordinates; all
simulations in this paper were performed in two dimensions with
cylindrical coordinates.

\subsubsection{Equation of State} 

CASTRO can follow an arbitrary number of isotopes or elements. The
atomic weights and amounts of these elements are used to calculate 
the mean molecular weight of the gas required by the equation of 
state.  We followed the elements from hydrogen through zinc, so that 
our elemental yields could be compared with observations of abundances 
in metal-poor stars.  {\Ni} was followed separately in order to 
calculate the energy deposited by radioactive decay of {\Ni} and \Co.

The equation of state in our simulations assumed complete ionization 
and included contributions from both radiation and ideal gas pressure:
\begin{equation}
        P= f(\rho,T) \, \frac{1}{3} a T^4 + \frac{k_{B} T \rho}{m_p \mu}
\end{equation}
\begin{equation}
       E = f(\rho, T)\, \frac{a T^4}{\rho} + 1.5\frac{k_{B} T}{m_p \mu},
\end{equation}
where $P$ is the pressure, $a$ is the radiation constant, $k_{B}$ is
Boltzmann's constant, $T$ is the temperature, $\rho$ is the density,
$m_p$ is the proton mass, $\mu$ is the mean molecular weight, and $E$
is the energy. The function $f(\rho, T)$ is a measure of the 
contribution of radiation pressure to the equation of state.  It is
1 in regions where radiation pressure is important, i.~e. where gas
is optically thick, and 0 in regions where radiation pressure is 
unimportant, i.~e. where gas is optically thin, with a smooth transition 
in between.  The function $f(\rho,T)$ takes the form

\[ f(\rho,T) = \left\{ \begin{array}{ll}
                           1                 & \mbox{if $\rho \ge 10^{-9}$ gm cm$^{3}$} \\
& \mbox{or $T \le T_{neg}$} \\
                             f(T)=e^{\frac{T_{neg}-T}{T_{neg}}}   & \mbox{if $\rho < 10^{-9}$ gm cm$^{3}$}\\
 & \mbox{and $T > T_{neg}$}\\
                          \end{array}
                  \right.\vspace{0.1in} \]

where $T_{neg}$ is the temperature at which contributions to the
pressure from radiation are negligable at 100 times less than that
contributed by ideal gas pressure:
\begin{equation}
  T_{neg}={\frac{3 k_b \rho }{100 m_p \mu a}}^{1/3}.
\end{equation}

Above this temperature, contributions to the pressure from radiation
will begin to dominate the equation of state, even though radiation
pressure is unimportant in optically thin regions.  Damping the
radiation component of the EOS in optically thin regions prevented it
from dominating in regions where it is not important.

\subsubsection{Radioactive Decay of \Ni}
Energy from the radioactive decay of  $^{56}$Ni to $^{56}$Fe was 
deposited locally at each mesh point.  The energy deposition rate
from the decay of \Ni \ to \Co \ was:
\begin{equation}
               dE_{\Ni} = \lambda_{\Ni} X_{\Ni}
               e^{-\lambda_{\Ni}t}q(\Ni).
\end{equation}
The decay rate of \Ni, $\lambda_{\Ni}$, is $1.315\times10^{-6}$
s$^{-1}$, and the amount of energy released per gram of decaying \Ni
\ is $q(\Ni)$, which we set to $2.96 \times 10^{16}$ erg g$^{-1}$ \citep{Woosley1988}.  
$X_{\Ni}$ is the mass fraction of \Ni \ in the cell.  The mass 
fraction of \Co \ at a given time can be expressed in terms of 
the mass fraction of initial \Ni \ by
\begin{equation}
               X_{\Co} =
               \frac{\lambda_{\Ni}}{\lambda_{\Co}-\lambda_{\Ni}} 
               X_{\Ni} (e^{-\lambda_{\Ni}t}-e^{-\lambda_{\Co}t}),
\end{equation}
so that the energy deposition rate from \Co \ is
\begin{equation}
               dE_{\Co} =
               \frac{\lambda_{\Ni}}{\lambda_{\Co}-\lambda_{\Ni}}
               X_{\Ni} (e^{-\lambda_{\Ni} t}-e^{-\lambda_{\Co} t}))
               \lambda_{\Co} q(\Co).
\end{equation}
We used a decay rate $\lambda_{\Co} = 1.042 \times10^{-7}$ 
s$^{-1}$ and an energy per gram of decaying \Co, $q(\Co)$, equal 
to $6.4 \times 10^{16}$ erg g$^{-1}$ \citep{Woosley1988}. 

\subsubsection{Gravity}

\code{CASTRO} supports several different approaches to solving 
for self-gravity. In the calculations presented here, we applied 
the monopole approximation for gravity.  A radial average of the 
density was taken from the two-dimensional grid to create a 
one-dimensional density profile.  This profile was then used to 
compute a one-dimensional gravitational potential which was then 
mapped back onto the two-dimensional grid. Since perturbations 
from spherical symmetry in the densities are very small, this 
approximation has a negligible effect on the final state of the 
calculation.  A comparison between the monopole gravity solver and 
the full multigrid solver can be found in \citet{Almgren2010}.  
The radial approximation to the gravitational potential has the 
advantage of being nearly as accurate as a multigrid solution
for these explosions, but is much faster.

Gravity from a point mass located at the origin was also included in
the gravitational potential.  The point mass represents the compact
remnant left behind by the SN explosion.  As infalling matter crosses
the zero-gradient inner boundary near the origin, it is added to this
point mass.

\subsubsection{\code{CASTRO} AMR}
\lSect{AMR} 

The adaptive mesh refinement algorithm in \code{CASTRO} uses a nested
hierarchy of logically-rectangular grids with simultaneous refinement
of the grids in both space and time.  The integration algorithm on the
grid hierarchy is a recursive procedure in which coarse grids are
advanced in time, fine grids are advanced multiple steps to reach the
same time as the coarse grids, and the data at different levels are
then synchronized.

There is a regridding step in which increasingly finer grids are recursively
embedded in coarse grids until the solution is sufficiently resolved.
\code{CASTRO} uses as its default refinement criteria the ``error
estimator'' of \citet{Lohner1987}, which is essentially the ratio of
the second derivative to the first derivative at the point at which
the error is evaluated.  Details of the implementation in various
geometries are discussed more fully in \citet{Almgren2010}. The result
is a dimensionless, bounded estimator, which allows arbitrary
variables to be used with preset error indicators.

We used density, pressure, velocity, and the abundances of \Ni, He, 
and O as our refinement variables.  We slightly modified the
refinement criteria for elemental abundance so that they were applied only
in regions where the abundance was greater than $10^{-3}$.  Regions
with abundances of an element lower than $10^{-3}$ were not marked for
refinement on the basis of elemental abundance alone.

We also modified the refinement criterion for density.  In order to
control extraneous refinement as the star expanded on the grid,
regions below a certain density were not marked for refinement below a
certain refinement level.  We chose a minimum ``mass'' at which
refinement could occur, which was the same for all the simulations.
The minimum density as a function of refinement level at which a cell
can be tagged for refinement is given by
\begin{equation}
\rho_{min,i} = 2^{2i}/A_o \times 10^{20} \: \mathrm{gm} \: \mathrm{cm}^{-3},
\end{equation}
where $\rho_{min,i}$ is the minimum density at which refinement can
occur at a given refinement level $i$ and $A_o$ is the area of the
cell at the $0th$ refinement level.  The factor of $10^{20}$ was
chosen because it kept refinement at the outer edge of the star to a
reasonable level, where it was resolved with approximately the same
number of cells as the outer edge of the helium shell, without losing
important detail in the RT instabilities.

\subsubsection{Initialization of Multidimensional Data}

In mapping the radial data from \code{KEPLER} onto the two-dimensional
axisymmetric grid in \code{CASTRO}, special care was taken to properly
resolve the key elements of the simulations: the shock, the elemental
shells, and the Fe core. In particular, both the Fe core and the O
shell were resolved with a minimum of 16 cells.  Doubling the
resolution produced no essential change to the solution.  Mixing
ceased at same time, and the final distribution of elements as a
function of mass was the same for simulations in which the Fe core was
resolved with 16 cell and 32 cells.  The one-dimensional mapping
results in explosions that are spherically symmetric in \code{CASTRO}:
low-order departures from spherical symmetry are not included in the
initial conditions, and do not develop in the course of the
simulation.. Our models therefore only capture asymmetries of mode
greater than $l=1$ or 2 in the explosions.

\section{Progenitor Models}
\lSect{models}

\begin{deluxetable}{lcr}
\tabletypesize{\scriptsize}
\tablecaption{Blast Models: Total Angular Momentum and Fraction
of Critical Angular Velocity \label{tab:rotation}}
\tablehead{
\colhead{model} & \colhead{$L_{tot}$ (gm cm$^{2}$s$^{-1}$)}\footnotemark[1]
                    & \colhead{\% v$_{\theta,crit}$}}\footnotemark[2]
\startdata
 z15-5   & 6.1407E+51  & 15.5 \\
 z25-5   & 6.1407E+51  &  6.2 \\
 z40-5   & 6.1407E+51  &  2.8 \\
	 &	       &      \\
 u15-5   & 6.1407E+51  & 11.3 \\
 u25-5   & 6.1407E+51  &  4.6 \\
 u40-5   & 6.1407E+51  &  2.1 \\
	 &	       &      \\
 z15-10  & 1.2281E+52  & 31.5 \\
 z25-10  & 1.2281E+52  & 12.4 \\
 z40-10  & 1.2281E+52  &  5.5 \\
	 &	       &      \\
 u15-10  & 1.2281E+52  & 23.3 \\
 u25-10  & 1.2281E+52  &  9.2 \\
 u40-10  & 1.2281E+52  &  4.1 \\
\enddata
\footnotetext[1]{Total angular momentum with which model was intialized}
\footnotetext[2]{Percent critical angular velocity of the star this amount of angular momentum represents}
\end{deluxetable}

To create the explosion profiles that were mapped into \code{CASTRO},
the one-dimensional Lagrangian stellar evolution code \code{KEPLER}
\citep{Weaver1978,Woosley2002} was first used to evolve the
progenitors through all stable stages of nuclear burning, up until
their iron cores became unstable to collapse.  At this point SN
explosions were artificially initiated by means of a piston at
constant Lagrangian mass coordinate that moved though the star with a
specified radial history.  The models presented in this paper used a
piston located at the radius where the entropy was equal to $4.0
k_B/$baryon, which corresponded roughly to the base of the oxygen
shell. The models were then evolved until all nuclear burning was
completed ($\sim$ 100 s after the start of the explosion) but prior to
the exit of the forward shock from the helium shell (and hence before
the formation of the reverse shock).  Energy generation was followed
with a 19-isotope network up to the point of oxygen depletion in the
core of the star and with a 128 isotope quasi-equilibrium network
thereafter. Rotation was included in the models, which had the effect
of increasing mixing in semiconvective regions of the star,
i.e. regions that are stable by the Schwarzchild but unstable by the
Ledoux criteria.

Our supernova models were taken from the survey of Heger (in prep).
We adopted progenitor masses of 15, 25, and 40 $\Msun$, which are
thought to be on the lower end of the mass scale for Population III
stars \citep[e.~g.][]{on07,yet07}. These masses were chosen in part
because they were considered in previous studies of zero-metallicity
stars \citep{Joggerst2009, Tominaga2009, Tominaga2007, Heger2008}.  We
examined both zero-metallicity and $Z=10^{-4} \Zsun$ SN progenitors.
These two metallicities correspond to Population III stars and to
stars on the cusp of the transition to Population II, where the gas
has enough metals to cool efficiently and form low mass stars that can
survive to the current epoch
\citep[e.~g.][]{mbh03,Frebel2007}. Important differences in stellar
evolution arise in these two metallicities; in particular, the $10^{
  -4}$ $\Zsun$ stars are less affected by rotation and are actually
bluer than their rotating $Z = 0$ counterparts.

We chose models with explosion energies of 0.6, 1.2, and 2.4 Bethe,
where 1 Bethe $ = 10^{51}$ ergs \citep[1.2 Bethe is near the observed median
explosion energy for core-collapse SNe ][]{Hamuy2003}.  The stars
were assumed to have no mass loss.  \citet{Nomoto2006} and
\citet{Umeda&Nomoto2005} proposed that the first generation of SNe may
have been ``hypernovae'' on the basis of observed abundances in the
most metal-poor stars. We instead decided to study models closer to
the more commonly observed explosion energies in the nearby
(admittedly solar-metallicity) universe.

Two fiducial rotation rates ``R'' were also considered: 5 and 10\% of
the critical angular velocity of a 20 \Msun \ solar-metallicity star,
or total angular momentum of 6.1407 $\times$ 10$^{51}$ or 1.2281
$\times$ 10$^{52}$ gm cm$^{2}$ s$^{-1}$, respectively. Models were
initialized with the same amount of angular momentum at a given
fiducial rotation rate.  These rates were 2 - 30\% of the critical
velocity for the stars in our survey, depending on their individual
mass and metallicity. This percentage is listed for each progenitor in
Table \ref{tab:rotation}. Although the percentage of the critical
velocity at a given ``R'' decreases with mass, this amount of angular
momentum is sufficient to drastically change the presupernova
structure of the model from a non-rotating counterpart.  The initial
angular momenta of our stars ensure that if they were of solar
metallicity, they would fall at typical to sub-typical rotation
rates. Solar metallicity massive stars typically rotate at about
$30\%$ of critical angular velocity. Rotation speeds for Population
III and II stars are not currently known, so in the absence of more
compelling data it is reasonable to adopt values typical of the local
universe.  Rapidly rotating stars would be more likely to explode with
a high degree of asymmetry, perhaps in the form of a jet, than more
slowly rotating stars.  The spherically symmetric explosions we have
constructed for this study are more consistent with the slowly
rotating progenitors we chose for our sample, though even slowly
rotating progenitors may explode asymmetrically (see
\citet{Murphy2009, Marek2009, Scheck2006} and references therein).

Rotation can dramatically change the structures of the progenitors, 
especially the zero-metallicity ones \citep{Ekstroem2008,Hirschi2007}.  
Rotationally induced mixing between the helium core and the base of 
the hydrogen shell introduces C, N, and O, which leads to CNO burning 
at the base of the hydrogen shell.  This dramatically increases the rate 
of energy production, puffing up the outer layer of the star.  It also 
changes the interior structure, resulting in a uniformly mixed 
helium-hydrogen layer outside the CO core for the zero-metallicity 
models. Including even a small amount of angular momentum in the models 
effectively turns compact blue zero-metallicity stars into red giants.  
This has a profound effect on their post-explosion hydrodynamics.  
Rotation produces less of an effect in stars with low metallicity

The nomenclature for our models is as follows.  Metallicity is denoted 
by either a ``u'' or a ``z'' at the beginning of the model name: ``u'' 
signifies $10^{-4}$ \Zsun \ while ``z'' means zero metallicity.  Next 
is the mass of the star: 15, 25, or 40 \Msun.  The explosion energy is 
denoted by either an ``B'', ``D'', or ``G'', corresponding to 0.6, 1.2, 
and 2.4 Bethe, respectively.  The rotation rate R is designated after the 
explosion energy by either a 5 or 10, the two fiducial rates defined 
above.  For example, Z25D-5 is the 25 \Msun, zero-metallicity model, 
exploded with 1.2 Bethe of energy and rotating at R$=5$.

\section{RESULTS}
\lSect{results}

The final composition of the ejecta of a core-collapse SN is
determined by the degree to which the layers surrounding the compact
remnant are mixed by the RT instability and by the amount and 
composition of material that accretes onto the remnant.  Mixing and 
fallback are determined by the presupernova structure of the star 
and by the details of its explosion.

\subsection{Initial Stellar Structure and the Onset of Rayleigh-Taylor 
Mixing }

At the end of its life, a massive star is composed of shells whose 
isotopic weight increases toward the Fe core in the center of the 
star.  These layers become unstable to RT mixing during the explosion. 
Heavier material from deeper layers is mixed outward and lighter 
material from shallower layers is mixed inward. A reverse shock forms 
when the outgoing shock encounters a region of increasing $\rho r^3$ 
\citep{Herant&Woosley1994,Woo95}.  When the shock encounters densities 
that fall less steeply than $r^{-3}$, i.e. a region of increasing 
$\rho r^3$, it decelerates.  The deceleration of the forward shock 
reverses the direction of the pressure gradient, 
which slows down the layers interior to the shock as well.  Shocked 
material piles up into a high density shell. The reverse shock forms 
at the inner boundary of the high-density shell of decelerated matter 
and propagates down into the star toward its center, slowing down 
the deeper, inner layers of the star \citep{Kifonidis2003}.  The
deceleration of the shock creates a steep pressure gradient in the
opposite direction to the gravitational and density gradients.  The 
pressure gradient may overwhelm the gravitational gradient, and in 
doing so trigger the formation of RT instabilities in the material.  
The RT instability grows until the reverse shock has passed by, at 
which point the material becomes stable again and the instabilities 
cease to grow exponentially.

\begin{figure*}
\centering
\plotone{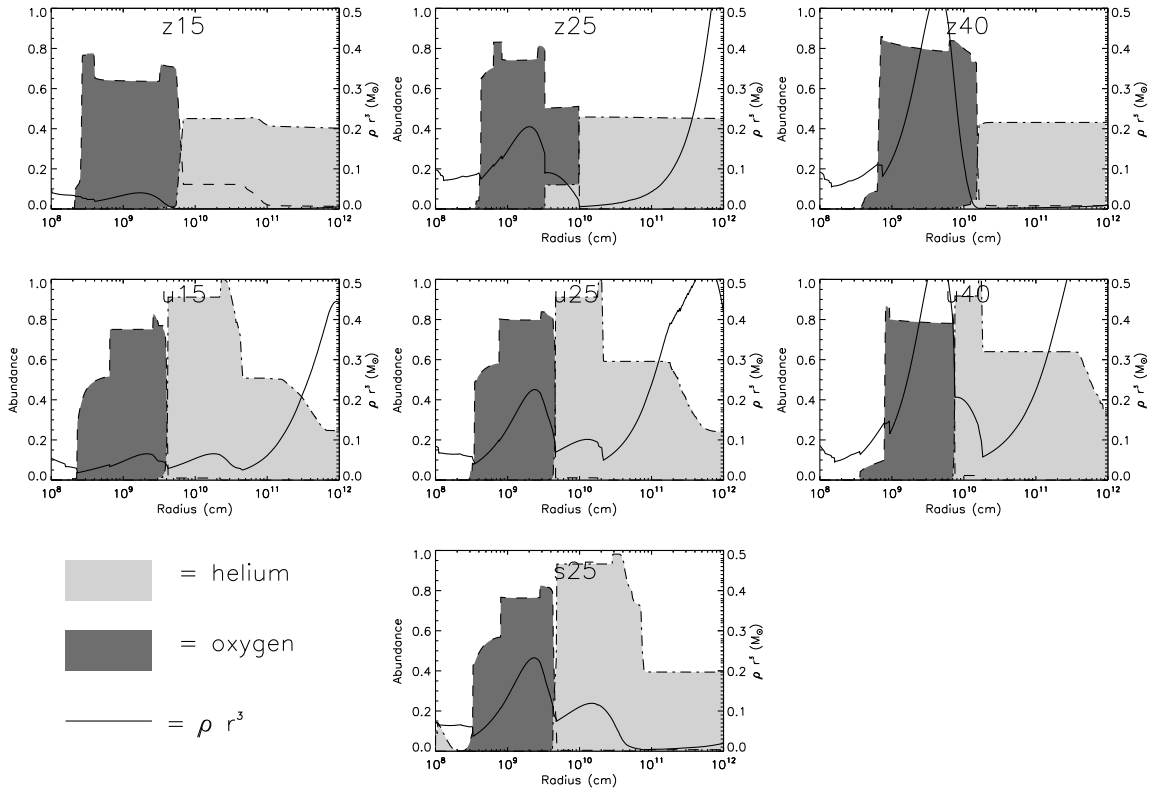}
\caption{Pre-supernova structure for $Z = 0$ (top row) and $Z=10^{-4}$
  \Zsun \ (middle row) used in this survey, along with the presupernova
  structure of a solar-metallicity star (bottom row) for comparison. The
  helium and oxygen shells are shown in light grey and dark grey,
  respectively, and the quantity $\rho r^3$ is shown as a solid line.
  When the forward shock encounters a region of increasing $\rho r^3$,
  a reverse shock forms, which leads to the formation of the R-T
  instability.  For the z-series models, this region is the boundary
  between the helium-hydrogen and oxygen shells.  For the $Z=10^{-4}$
  $\Zsun$ stars, the boundaries between both the helium and hydrogen 
  shells and the oxygen and helium shells are unstable.}
\label{fig:rhor3}
\end{figure*}

The presupernova structure of the star gives some indication of where
the reverse shock, and hence the RT instability, is likely to form.
Figure \ref{fig:rhor3} shows $\rho r^3$ for the R=10 progenitors in
this study, scaled to show the location and shape of the helium and
oxygen shells as well.  We will show the final mixed structures of
these stars in section \Sect{RTmixing}.  The structure of the
progenitor for R=5 is qualitatively similar.  Shown for comparison is
a 25 \Msun\, $Z=$\Zsun\ star from \citet{Herant&Woosley1994}.  The
$Z=$\Zsun\ and $Z$=$10^{-4}$ \Zsun \ models are similar in that $\rho
r^3$ does not increase dramatically until the hydrogen-helium shell
boundary.  The reverse shock that drives RT mixing will form at this
boundary. The $Z$ = 0 models are different.  In these models, the
shell boost at the base of the hydrogen layer has led to so much
convection that a distinct helium shell no longer exists, so although
the $Z=0$ \Zsun \ models are as large in radial extent as the
$Z=\Zsun$ models, the former lacks the helium shell present in the
latter.  Instead, helium and hydrogen are nearly uniformly mixed
beyond the oxygen shell.  The quantity $\rho r^3$ increases sharply at
the boundary between the oxygen shell and the helium-hydrogen shell,
indicating that the reverse shock will form, and RT mixing will
commence, at this boundary deeper within the star.

\subsection{Effects of Rotation on Mixing}
\lSect{effectsofrotationonmixing}

\begin{figure}
\centering
\plotone{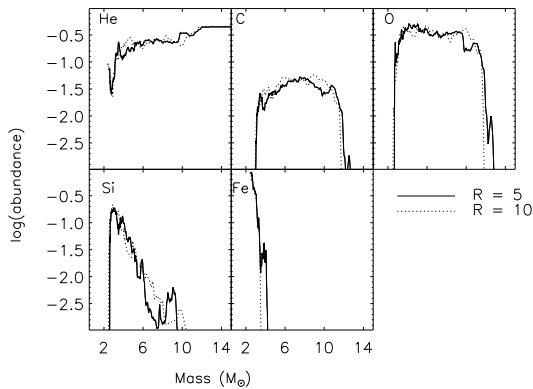}
\caption{Comparison of elemental abundance as a function of mass
  coordinate for star z15G.  The solid line is the model z15G-5 (see
  Table \ref{tab:rotation}); the dotted line corresponds to model
  z15G-10. No substantial difference in element distribution was
  found between the different rotation speeds for any stellar
  model. This model is representative of the average differences
  between stars at R=5 and R=10.}
\label{fig:rotcomp}
\end{figure}

Figure \ref{fig:rotcomp} shows the abundance distribution of
individual elements as a function of mass for models z15G-5 and
z15G-10.  Although they are not identical, the distributions are quite
similar for the two models.  The other R=10 SN models in our sample
displayed a similar if not higher degree of congruence with their R=5
counterparts.  If the star rotates, the degree of rotation has little
effect on the final distribution of elements in the explosions, at
least for the rotation rates explored in our survey.  The reason for
this is that once a $Z=0$ star acquires even a small amount of
rotation and becomes a red giant rather than a compact blue star
\citep{Joggerst2009}, the structure of the red giant is relatively
insensitive to the degree of rotation.  Rotating $Z=0$ stars are about
an order of magnitude larger in radial extent than non-rotating $Z=0$
stars, though our 25 \Msun \ rotating models are closer in size to
their non-rotating counterparts than 15 and 40 \Msun \ models.
Consequently, as the ejecta propagates through the star it encounters
nearly the same gas profile whether it is an R=5 or an R=10 star and
becomes similarly mixed.  Mixing within the R series in $Z = 10^{-4}
\Zsun$ stars is similar because the introduction of a modest amount of
rotation has little effect on the structure of progenitor in these low
metallicity models.  Because models of either metallicity were so
similar for R=5 and R=10, we will focus on the R=10 models for the
remainder of this paper.  However, because the structures of red
giants are so distinct from compact blue stars, mixing and fallback
vary dramatically with metallicity, as we discuss next.

\subsection{RT Mixing in Models by Mass, Metallicity and Energy}
\lSect{RTmixing}

\begin{figure*}
\centering
\plotone{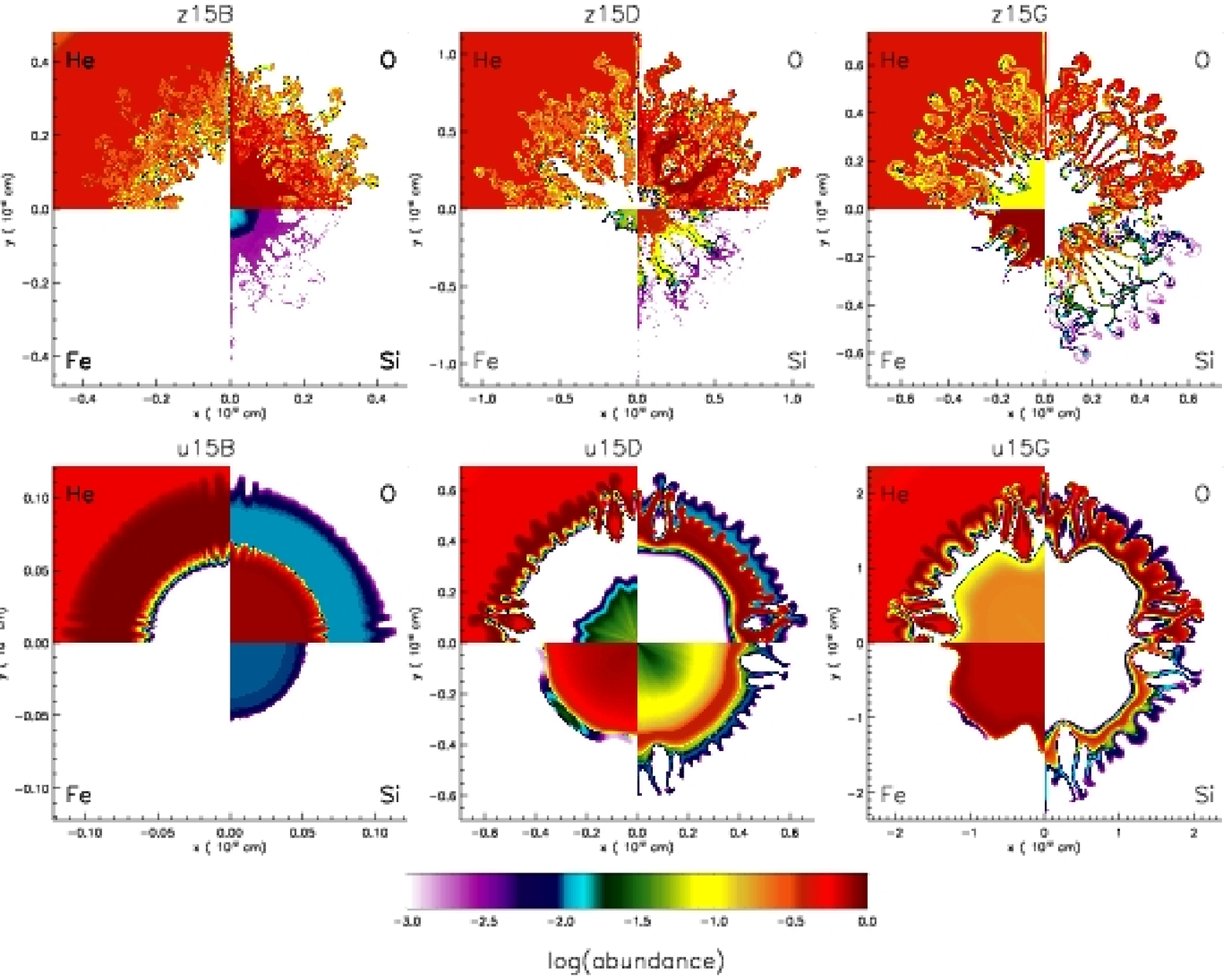}
\caption{Mass abundance of He, O, Si, and Fe in $Z = 0$ (top) and
  $10^{-4}$ \Zsun \ (bottom) 15 \Msun \ stars after RT-driven mixing
  in the model has ceased.  $Z= 0$ stars, which die as large red
  giants, show much more mixing than $Z=10^{-4}$ \Zsun \ stars, which
  die as smaller blue giants.  The snapshots are of the simulation at
  $3.3 \times 10^{5}$ s, $4.4 \times 10^5$ s, and $4.0 \times 10^5$ s
  for z15B, z15D, and z15G, respectively, and similarly, at $7.9
  \times 10^4$ s, $6.1 \times 10^4$ s, and $5.0 \times 10^4$ s for
  u15B, u15D, and u15G.  Mixing increases with explosion energy, which
  is 0.6, 1.2, 2.4 Bethe from left to right across the panels. The
  jetting along the y- and x-axes, a numerical artifact, causes the
  departures from spherical symmetry in the u-series models but does
  not substantially affect the conclusions in this paper.}
\label{fig:pic15}
\end{figure*}

\begin{figure*}
\centering
\plotone{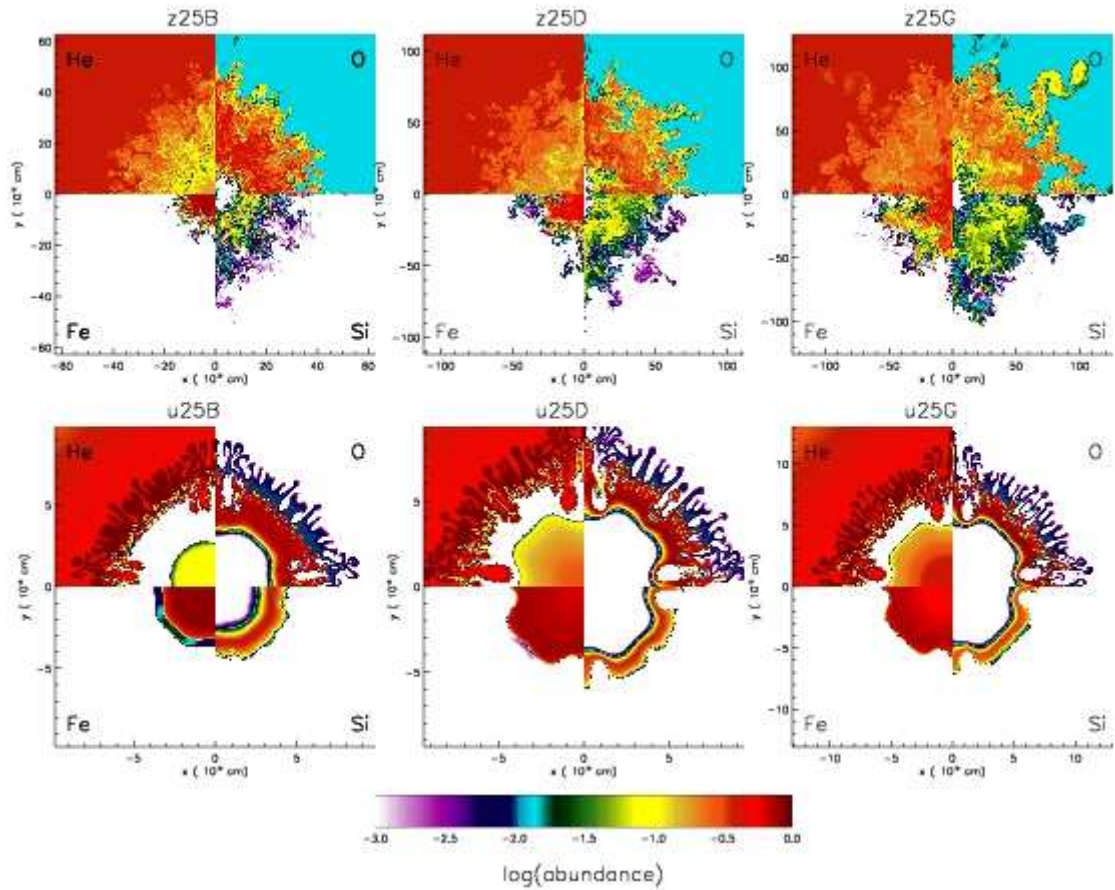}
\caption{Mass abundance of He, O, Si, and Fe in $Z = 0$ (top) and
  $10^{-4}$ \Zsun \ (bottom) 25 \Msun \ stars after the end of
  RT-driven mixing.  The snapshots are of the simulation at $3.1
  \times 10^4$ s, $6.3 \times 10^4$ s, and $2.7 \times 10^4$ s for
  z25B, z25D, and z25G, and $1.4 \times 10^4$ s, $5.3 \times 10^4$ s,
  and $1.2 \times 10^5$ s for models u25B, u25D, and u25G,
  respectively.  Red $Z = 0$ stars again show much more mixing than
  blue $Z=10^{-4}$ \Zsun \ stars, although it is not as extreme as in
  the 15 \Msun \ models, in which the difference in outer radius
  between the z- and u-series progenitors was greater. Mixing again
  rises with explosion energy, which is 0.6, 1.2, 2.4 Bethe from left to
  right across the panels. Spurious jetting is also visible along the
  y- and x-axes in the u-series models.  Like the 15 \Msun \ stars
  shown in Figure \ref{fig:pic15}, both mixing and the amplitudes of
  the RT instabilities clearly increase with explosion energy at both
  metallicities.}
\label{fig:pic25}
\end{figure*}

\begin{figure*}
\centering
\plotone{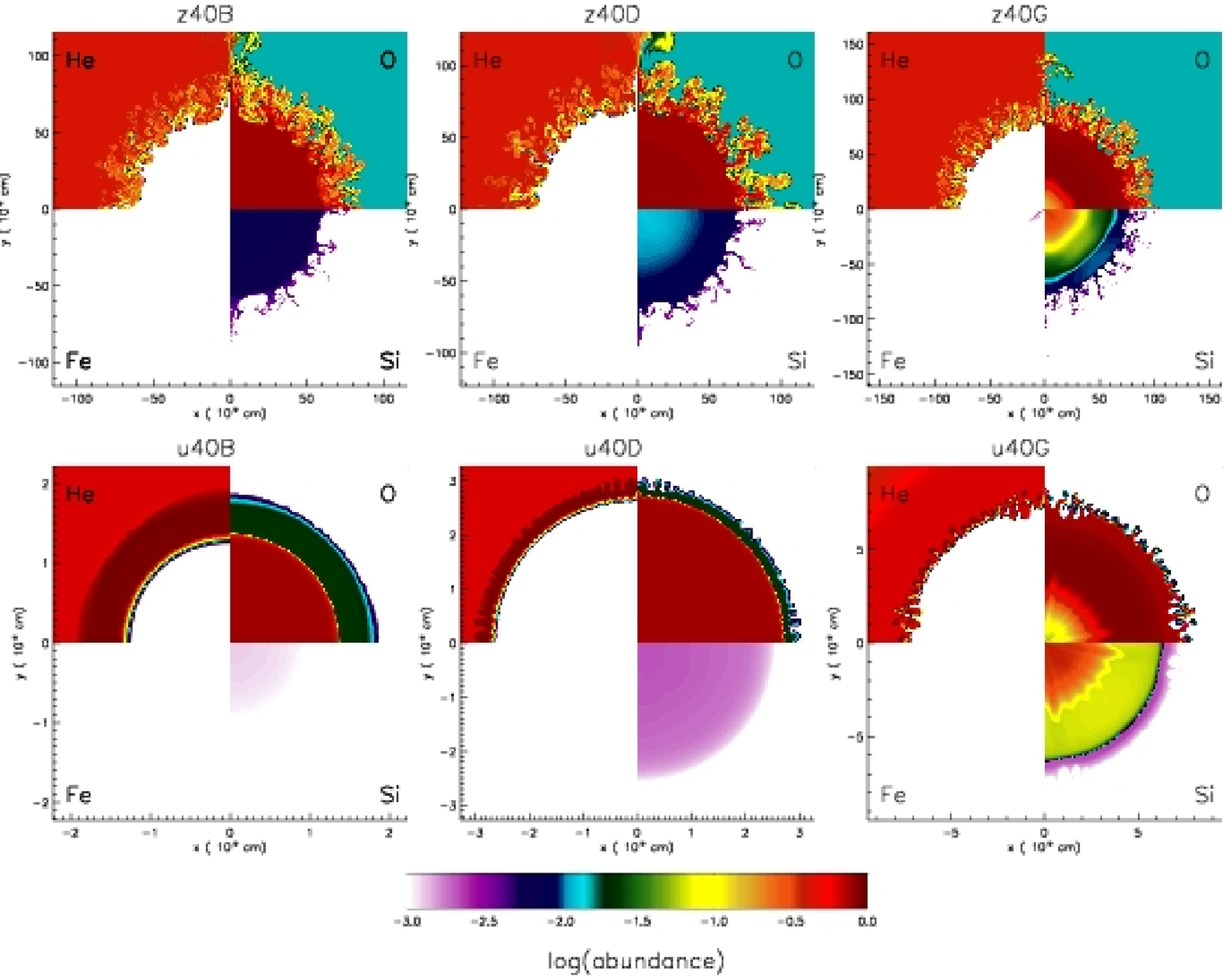}
\caption{Mass abundance of He, O, Si, and Fe for $Z = 0$ (top) and
  $10^{-4} \Zsun \ (bottom) $ 40 \Msun \ stars at $4.3 \times 10^5$ s,
  $3.5 \times 10^5$ s, and $2.9 \times 10^5$ s for models z40B, z40D,
  and z40G, and $2.5 \times 10^4$ s, $2.6 \times 10^4$ s, and $4.9
  \times 10^4$ s for models u40B, 740D, and u40G, respectively.  The
  same trends in mixing with mass, metallicity, and explosion energy
  as in the 15 and 25 \Msun \ stars are evident in these models.}
\label{fig:pic40}
\end{figure*}

The radius of the outer envelope of stars of a given mass is quite
different for the two metallicities. The $Z = 0$ models are much
redder than the $Z=10^{-4}$ \Zsun \ models because their outer
envelopes are larger.  This is due to the hydrogen shell boost that
dramatically increased the rate of hydrogen burning at the base of the
hydrogen shell, making the outer envelope of the star entirely
convective as it left the main sequence.  The outer radii of the $Z =
0$ stars ranged from $6.7 \times 10^{12}$ to $1.32 \times 10^{14}$ cm
for the 15 to 40 $\Msun$ stars, respectively.  The outer radii of the
$Z=10^{-4}$ \Zsun\ stars were about an order of magnitude smaller,
ranging from $2.4 \times 10^{12}$ cm for the 25 \Msun \ star to $4.18
\times 10^{12}$ for the 40 \Msun \ star.  Since the time it takes the
reverse shock to propagate through the layers of the star sets the
timescale on which RT mixing can occur, we expect that RT
instabilities will have more time to develop, and greater mixing will
occur, in the more ``puffed up'' $Z=0$ stars.  Stronger explosions
will induce more mixing than weaker ones.  The strength of the reverse
shock, the degree to which the pressure gradient is reversed, and
therefore the violence of the RT instabilities depends on the strength
of the initial shock and hence on the energy of the initial explosion.

We evolved the initial \code{KEPLER} supernova models in \code{CASTRO}
past the formation of the reverse shock and RT instabilities, until RT
mixing ceased and the SN remnant was expanding nearly homologously.
We found that the time at which RT mixing halted, and the degree to
which it mixed together the isotopic shells of the model, indeed
varies with mass, metallicity, and explosion energy.  Results for the
18 R=10 models are presented by mass in Figures \ref{fig:pic15},
\ref{fig:pic25}, and \ref{fig:pic40}.

\subsubsection{15 \Msun \ Models}

In Figure \ref{fig:pic15} we show the mass abundance of helium,
oxygen, silicon, and iron in 15 \Msun \ stars of 0.6 (B, at left) 1.2
(D, center) and 2.4 (G, at right) Bethe explosions after the end of RT
mixing.  The snapshots are of the simulation at $3.3 \times 10^{5}$ s,
$4.4 \times 10^5$ s, and $4.0 \times 10^5$ s for z15B, z15D, and z15G,
respectively, and similarly, at $7.9 \times 10^4$ s,$6.1 \times 10^4$
s, and $5.0 \times 10^4$ s for u15B, u15D, and u15G.  Beyond these
times, the distribution of elements as a function of mass does not
change.  The zero-metallicity models (top panels) show a higher degree
of mixing than their $Z=10^{-4}$ counterparts, as expected.  As
explosion energy increases (from left to right in Figure
\ref{fig:pic15}) the degree to which the inner layers of the star are
mixed also increases.  This is most apparent in the u-series models.
In model u15B-10, the RT instability does not fully develop, leaving
only a hint of ``fingers'' at the H-He and He-O shell boundaries. In
model u15D-10, where the explosion energy has doubled, RT mixing has
penetrated as far as the silicon shell, and silicon fingers extend
well into the oxygen shell.  Iron, however, remains unmixed, even in
model u15G-10.  The $Z = 0$ models show nearly complete mixing,
however.  Fe from the inner core is mixed out to at least as far as
the oxygen shell, and the RT instability has become fully non-linear.
The original fingerlike morphology has been significantly blurred over
the many mixing timescales over which the instability has had time to
develop.

\subsubsection{25 \Msun \ Models}

We show mixing in the 25 \Msun \ models in Figure \ref{fig:pic25}.
Here, as at 15 \Msun, RT mixing has had a longer time to develop in
the z-series than in the u-series models. The instabilities have
evolved to the point that the initial fingers have been nearly erased
in the Z series.  The snapshots show the distribution of elements at
$3.1 \times 10^4$ s, $6.3 \times 10^4$ s, and $2.7 \times 10^4$ s for
z25B, z25D, and z25G, and $1.4 \times 10^4$ s,$5.3 \times 10^4$ s, and
$1.2 \times 10^5$ s for models u25B, u25D, and u25G, respectively.
The center of the star has become mixed, and iron has reached the
outer layers of the star.  The initial radii of the 25 \Msun \
Population III and Population II models differ only by a factor of 3
or so, less than the difference in initial radii between 15 or 40
\Msun \ stars of different metallicity.  This is why the amount of
mixing seen in the 25 \Msun \ models is more similar than it is for
stars at 15 and 40 \Msun.

While the figures do not show it clearly because they have been
cropped to show just the mixing in the star, ``jetting'', or excessive
flow along an axis, is present along the y-axis for the z-series
models and is more pronounced in the higher explosion energy models (D
and G) than in B. These jets extend as far as $2 \times 10^{14}$ cm,
or twice the radius of the mixed region, for the z-series models.  The
u-series models again exhibit more mixing at the higher explosion
energies, although the effect is not as pronounced as in the 15 \Msun
\ models.  The deviation from spherical symmetry in the inner, unmixed
shells of silicon and oxygen is a numerical artifact of the
axisymmetric geometry, which is discussed in greater detail below in
\Sect{numericalartifacts}.  The artifact becomes more prominent with
explosion energy.

\subsubsection{40 \Msun \ Models}

The 40 \Msun \ mass models are shown in Figure \ref{fig:pic40} at $4.3
\times 10^5$ s, $3.5 \times 10^5$ s, and $2.9 \times 10^5$ s for
models z40B, z40D, and z40G, and $2.5 \times 10^4$ s, $2.6 \times
10^4$ s, and $4.9 \times 10^4$ s for models u40B, 740D, and u40G,
respectively.  These explosions experience more fallback than the
lower mass models, especially at explosion energies of 0.6 Bethe,
where they leave behind remnants of 8.1 and 9.1 \Msun \ for z40B and
u40B, respectively.  Mixing also occurs in less time.  Even in the
red giant $Z = 0$ models, the Fe core has completely fallen back onto
the remnant at the center of the simulation by the time RT mixing
begins in all but the most energetic explosions, so these models
produce virtually no iron. The variation in mixing with explosion
energy is also more apparent in the 40 \Msun \ u-series models than in
the 15 and 25 \Msun \ u-series runs.

\subsubsection{Variation of Mixing with $E_{ex}, M, Z$}

\begin{figure*}
\centering \plotone{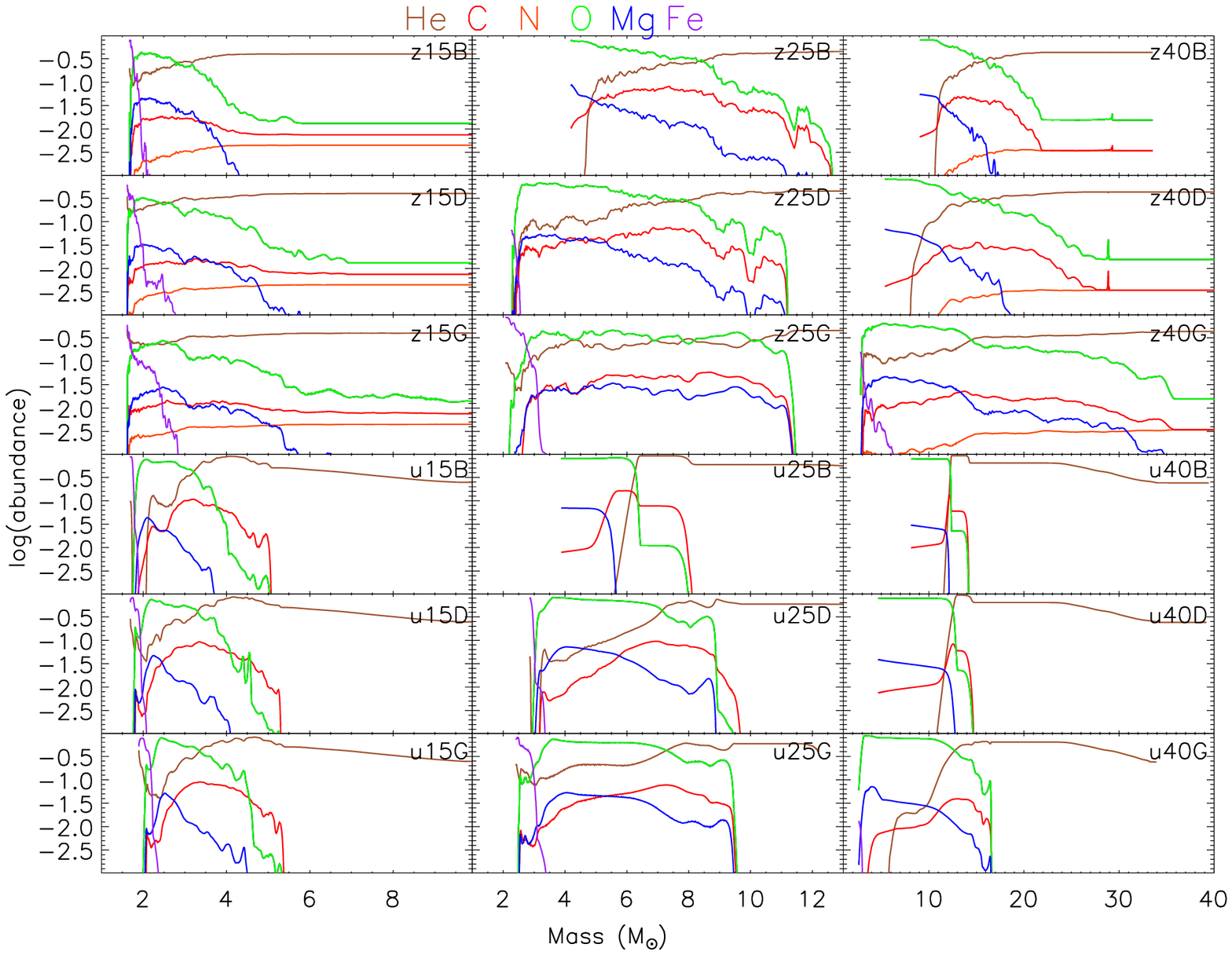}
\caption{Distribution of elemental abundances as a function of mass.
  Stellar mass increases from left to right; the top three panels are 
  $Z = 0$ stars; the bottom three rows are $Z=10^{-4}$ \Zsun \ stars.
  Explosion energy increases from top to bottom for each metallicity.  
  Individual elements are coded by color.  Zero-metallicity stars in 
  general show more mixing than the slightly enriched stars.  Higher 
  explosion energy SNe exhibit more mixing and less fallback, and lower 
  mass SNe have more mixing and less fallback.}
\label{fig:mpmp}
\end{figure*}

The final appearance of the SN ejecta at a given mass does not vary
qualitatively with rotation rate, but mass, metallicity, and explosion
energy do govern the final composition of the ejecta and the mass of
the remnant.  Figures \ref{fig:pic15}, \ref{fig:pic25}, and
\ref{fig:pic40} give a good qualitative picture of the morphology of
mixing in the explosion models; in Figure \ref{fig:mpmp} we quantify
the propagation of He, C, N, O, Mg, and Fe in the ejecta as a function
of mass coordinate for the 18 R=10 models after RT mixing has ceased.
Several trends are apparent.  Lower-mass models show less fallback and
more mixing in the internal layers than higher-mass models.  The
z-series SNe have far more mixing than u-series SNe.  SNe with higher
explosion energies exhibit more mixing and less fallback than SNe with
lower explosion energies.  In particular, the B series SNe with
sub-normal explosion energies, 0.6 Bethe instead of the canonical 1.2
Bethe, eject almost no iron with the exception of model z15B.

The z-series models all show more mixing than their u-series
counterparts.  The 25 \Msun \ models show the most mixing of the
models in the u-series, while the 40 \Msun \ u-series runs show the
smallest degree of mixing.  All the 40 \Msun \ models experience a
great deal of fallback, but the u-series models show the most because
they are more compact.  The higher explosion energy models exhibit
less fallback.

\subsubsection{Comparison with Kepler Estimations of Mixing}
\begin{figure*}
\centering
\plotone{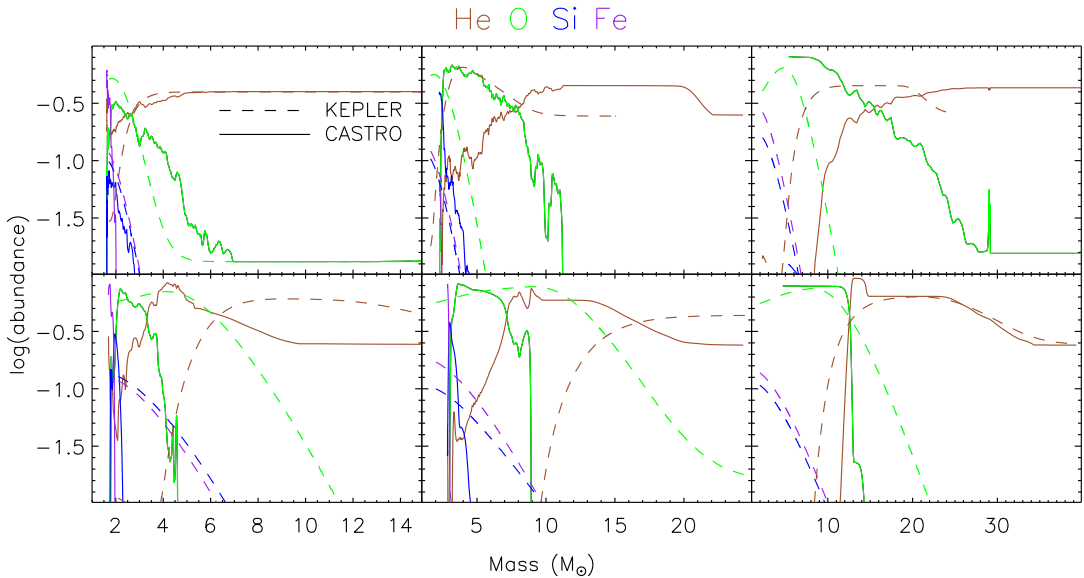}
\caption{Distribution of He, O, Si, and Fe for $Z = 0$ (top) and
  $10^{-4} \Zsun \ (bottom) $ 15 \Msun (left), 25 \Msun (center), and
  40 \Msun (right) for $1.2 \times 10^{51}$ erg explosions.
  \code{CASTRO} 2-d simulation results (solid lines) are compared with
  \code{Kepler} approximations of mixing (dashed lines). Kepler
  underestimates mixing of oxygen in the Z-series models, and
  overestimates mixing of elements heavier than oxygen. Kepler
  overstimates mixing for all elements as compared to the results of
  2-d simulations of U-series models.}
\label{fig:compmix}
\end{figure*}

The large one dimensional surveys of SNe derive final estimates of
elemental yields by artificially mixing the layers of the SN after
explosive nucleosynthesis is complete.  Surveys employing the
\code{Kepler} code estimate mixing by passing a running boxcar average
of width (in mass coordinate) $W$ through the star, where $W$ is
$10\%$ the mass of the helium core. That is, the abundances at points
that fell within a bin of width $W$ were averaged together and set to
this average, the bin was moved forward by one point, and the process
repeated, moving outward through the star.  This is done 4 times,
artificially mixing the mass shells.  In Fig \ref{fig:compmix} we
compare \code{KEPLER} estimations of mixing with our two dimensioanl
\code{CASTRO} results.  In our two-dimensional \code{CASTRO}
simulations, we find that some elemental shells are more mixed than
others.  The RT instability typically forms at the He-H or O-He
boundary and advances inward.  This results in the helium and oxygen
layers being more mixed than in \code{KEPLER} and the iron, and
sometimes silicon, layers being less mixed than the \code{KEPLER}
estimations for the Z-series models.  Our compact U-series models show
less mixing in all elements than in \code{KEPLER}.

\subsubsection{Numerical Artifacts \& Model Limitations}
\lSect{numericalartifacts} 

Numerical artifacts arising from the mesh geometry are most prominent 
in the higher explosion energy, u-series models, but they are present
in all the runs. Jetting is visible along both axes in the z-series 
models in Figures \ref{fig:pic15}, \ref{fig:pic25}, and \ref{fig:pic40}, 
but is more pronounced along the y-axis.  This is a well-known artifact 
that appears on axisymmetric coordinate grids: \code{PROMETHEUS} and 
\code{FLASH} \citep{Fryxell1991,Calder2002} also manifest it.  The 
deviation from spherical symmetry (most clearly visible in models 
u15D-10 and u15G-10) is also a result of the tendency of the shock to 
be most exaggerated in the direction in which it is moving, this time 
with respect to the reverse shock.  Ripples in the shells appear along 
both the x- and y-axes.

These anomalies do not change the essential results of our simulations.  
The z-series models would show a greater degree of mixing even without 
the presence of a jet directed either outward or inward.  In addition, 
because of the axisymmetric geometry, the mass lying along the y-axis 
is only $1/X$ (where $X$ is the distance along the x-axis) times the 
mass at a similar radius from the origin, so the jet contains little 
mass.  Additional mixing in the form of material pushed out farther 
than it should be along this axis was removed from the averages 
simply by excluding a $5^{\circ}$ cone of material around the axis.  
Fallback is not affected by jetting in our simulations.

In addition to these minor numerical effects, there are some inherent
limitations to our models.  As discussed in \Sect{numericalartifacts},
these explosions are spherical, and capture initial perturbations of
only much higher order than $l=1$ or 2.  Jet-driven explosions, as
well as those with asymmetries of order $l=1$ or 2, may be common in
the early universe, especially in rapidly rotating stars.  Such
explosions may have very different nucleosynthetic patterns
\citep{Tominaga2009}. By examining only more slowly rotating stars, it
is likely that the mapping of one-dimensional profiles onto
two-dimensional grids in \code{CASTRO} is not excluding serious
low-order perturbations, but not enough is known about the explosion
mechanisms of core-collapse SNe, let alone Population III and II SNe,
to be certain.

Simulations of spherical explosions in core collapse SNe, especially
those of blue progenitors like the u-series $Z=10^{-4}$ \Zsun \ SNe
presented here, have historically \textit{underestimated} mixing and
final elemental velocities \citep{Herant&Woosley1994, Fryxell1991,
  Kifonidis2003, Kifonidis2006}.  Previous studies that attempted to
match observations of SN 1987A (which died as a compact blue giant)
with techniques similar to those in our study were largely
unsucessful, finding less mixing and lower velocities for the heavier
elements than were observed.  \citet{Kifonidis2006} came closest to
reproduxing observations of SN1987A by adding additional energy to the
base of the explosion in two dimensions in the first few seconds after
the explosion.  We do not employ this technique in this paper.  Solar
metallicity models of red supergiants, like our $Z = 0$ models, come
closer to matching observations of similar red SNe
\citep{Herant&Woosley1994}.  It is possible that more mixing would
occur if the details of the first few seconds of the SN explosion were
known with greater accuracy. Our mixing estimates should therefore be
taken as lower bounds, though our mixing estimates for the red $Z=0$
\Zsun \ models in this paper are likely more accurate than for their
blue $Z=10^{-4}$ \Zsun \ counterparts.

\section{Discussion}
\lSect{discussion}

\subsection{Comparison with Metal-Poor Stars}
\begin{figure*}
\centering
\plotone{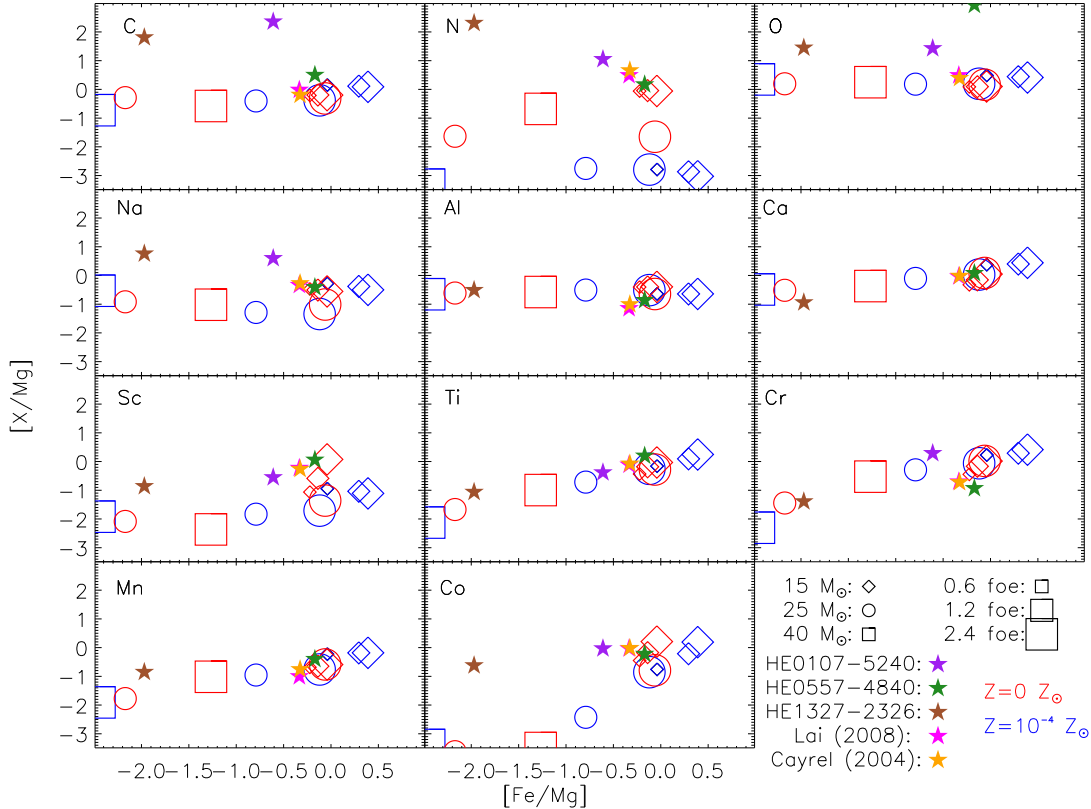}
\caption{Comparison of elemental yield ratios from SN models with 
  those observed in the 3 most metal-poor stars and with average 
  abundances for EMP stars from the surveys of \citet{Lai2008} and 
  \citet{Cayrel2004}. Observations are denoted by a star symbol; 
  models are denoted by open symbols where the size of the symbol 
  varies with explosion energy, shape with mass, and metallicity 
  with color.  None of our stellar yields fit the high [C+O/Fe] of 
  HE1327-2326, nor are we able to closely reproduce the abundance 
  pattern of HE0107-5240 for elements heavier than Na.  The z15D 
  and z15 yields are good fits to HE0557-4840.}
\label{fig:hmpscat}
\end{figure*}

Comparison of our theoretical yields to direct observations of $Z = 0$
and $Z=10^{-4}$ \Zsun \ SNe is not currently possible, as these
objects do not exist in the nearby universe.  However, the fossilized
yields of such explosions may be detectable in the abundance patterns
of extremely metal deficient halo stars.  The three most iron-poor
stars, HE0557-4840 \citep{Norris2007}, HE0107-5240
\citep{Christlieb2004}, and HE1327-2326 \citep{Aoki2006}, are thought
to have been enriched by at most one or a few SNe, while EMP halo
stars with $-4 < $[Fe/H]$ < -3$ are believed to be imprinted by
elements from a wider sample of early SNe.  In Figure
\ref{fig:hmpscat} we show our SN yields together with observed
abundances in the three HMP stars and the average of the abundances of
the larger sample of EMP stars in the \citet{Lai2008} and
\citet{Cayrel2004} surveys, which contain 15 and 22 stars,
respectively.

We calculate our theoretical yields by assuming the ejected material
is completely mixed, as is usual for this kind of calculation.  All
short-lived isotopes (\Ni \ for example) are reported as their
eventual decay products.

Figure \ref{fig:hmpscat} shows that our SN models do not reproduce the
high CNO abundances present in HE 01327-2326 and HE 0107-5240 when the
ratio of Mg to Fe is held fixed.  With its lower ratios of C and N to
Fe, HE 0557-4840 is a better match to some of our chemical yields, and
is also closer to the abundances in the \citet{Lai2008} and
\citet{Cayrel2004} data.  It is also clear from the diagram that some
models cannot produce the [Fe/Mg] values of any of the
observations. All but the most energetic of the 40 \Msun \ SNe yield
an [Fe/Mg] ratio too small to appear in Figure \ref{fig:hmpscat}.  The
25 \Msun \ explosion with energies less than 2.4 Bethe show little
overlap with either the HMP or EMP data, either.  The u-series 15
\Msun \ models produce a [Mg/Fe] ratio that is too high.

\subsubsection{HMP Stars}
\begin{figure*}
\centering
\plotone{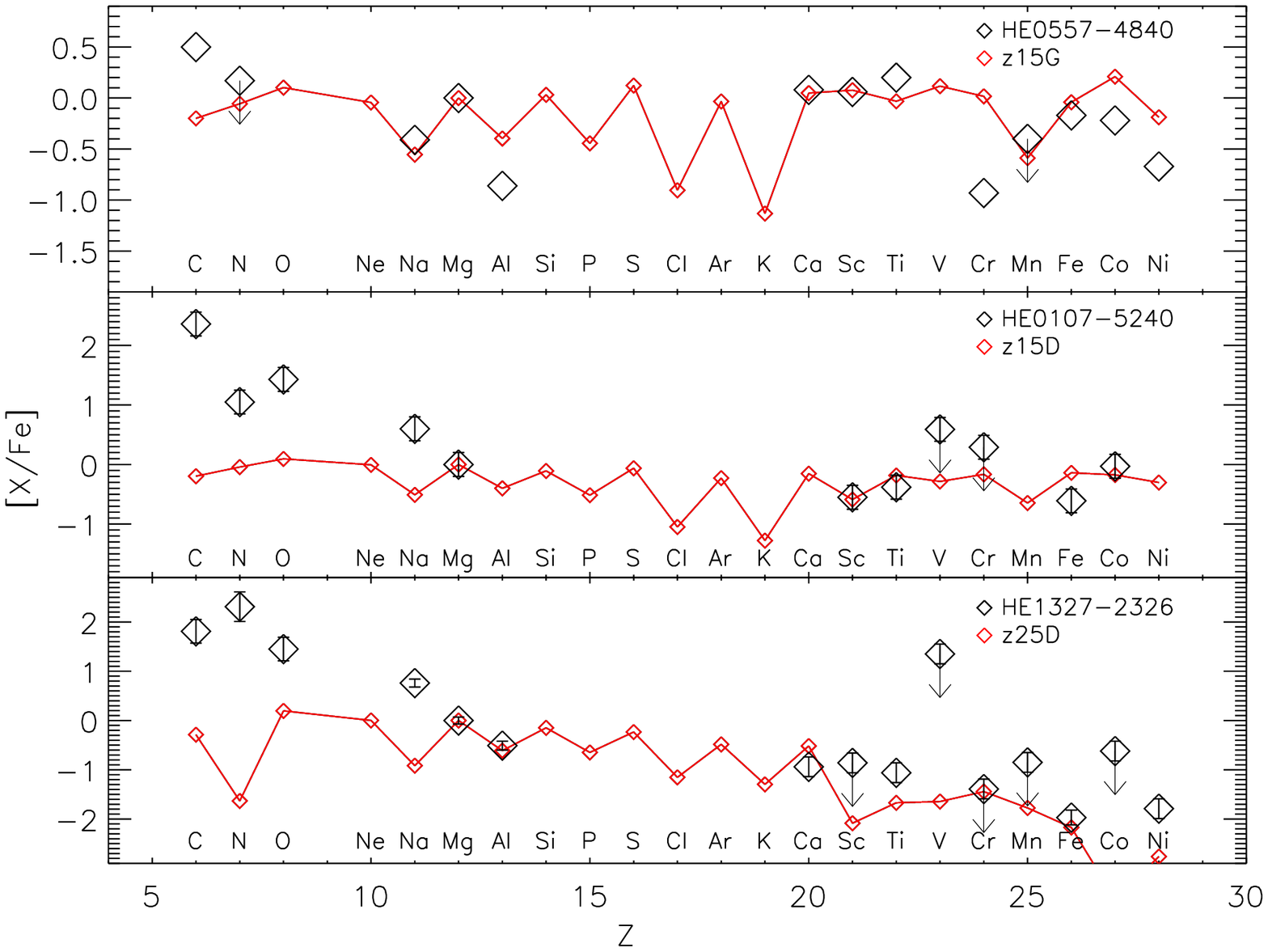}
\caption{Best-fit yields compared to observations of abundance
  patterns in the three most metal-poor stars.  The abundances
  in HE0557-4840 are well-reproduced by the yields from model z15G.  
  Abundances of elements heavier than Na are modelled well by z15D 
  for HE0107-5240 and bracketed by models z15D and z25D for 
  HE1327-2326, but these models do not produce the high C,N, O, and
  Na abundances observed in the most metal-poor stars. Enrichment by
  a companion AGB star might explain these excesses.}
\label{fig:hmpcomp}
\end{figure*}

While HE 0107-5240 and HE 1327-2326 show higher ratios of C,N, O and
Na to Fe than our models, the yield from one model, z15G, does fit the
abundance pattern of halo star HE 0557-4840 with the exception of
carbon.  Oxygen is an upper limit, and [C+O/Fe] in this star is far
lower than in the other two HMP stars.  The other two HMP stars, HE
0107-5240 and HE 1327-2326 show much higher [C+0/Fe] and [C+0/Mg]
ratios that our model yields do not replicate.  We compare yields from
the explosions in our survey that most closely reproduce [Mg/Fe] for
these two HMP stars with their observed abundance ratios in Figure
\ref{fig:hmpcomp}.  Model z15G provides a very close fit to the
abundance patterns in HE0557-4840. The only elements not in agreement
are carbon, where our model underpredicts [C/Mg] by about 0.5 dex, and
chromium, where our model underpredicts [Cr/Mg] by a little more than
1 dex.  \citet{Norris2007} compare their observed abundances for this
star to theoretical yields from the ``mixing and fallback'' models of
\citet{Umeda&Nomoto2003} and \citet{Iwamoto2005}, which overpredict
      [Cr/Fe]. The hypernova models of \citet{Umeda&Nomoto2005} and
      \citet{Nomoto2006} yield a [Co/Fe] ratio of 0.5, which is
      considerably higher than the observed value -0.3 for this star,
      or the value of -0.2 predicted by model z15G.

We also note that the non-rotating spherical explosion models in 
\citet{Joggerst2009} also fail to reproduce the abundance patterns 
found in metal-poor stars.  In particular, far less nitrogen is 
formed in their models than observed or produced in the rotating 
models.  This, not surprisingly, suggests that Population III stars 
were indeed rotating, but at the fairly low speeds modeled here the 
exact rate seems to be unimportant.

Reconciling our models to the abundance patterns in HE0107-5240 and
HE1327-2326 is more problematic.  Enrichment by a companion AGB star
has not been ruled out for either star, and would result in abundances
of C, N, O, and Na well above that with which the stars were initially
born.  This has been proposed as an explanation for the abundances in
HE0107-5240 \citep{Christlieb2004} and HE1327-2326 \citep{Aoki2006},
although \citet{Frebel2008} presents evidence that HE1327-2326 is not
a member of a binary system and thus could not have been enriched by
an AGB companion. While the mixing and fallback models of \citet{
  Iwamoto2005} account for the abundance patterns in these stars, they
are still heavily tuned one-dimensional parameterizations, not
multidimensional calculations with realistic physics.  The jet-driven
explosions of a 40 \Msun \ Population III star by \citet{Tominaga2009}
do reproduce the features of the earlier mixing and fallback models,
providing a mechanism for the mixing-fallback parameterization.  Their
angle averaged yields match the abundances observed in HE1327-2326 and
HE0107-5240--in particular, they produce the high [CNO/Fe] ratios that
our models do not.

Our models never show [C/Mg] and [O/Mg] ratios greater than 0.25 and
0.5, respectively. HE0107-5240 has [C/Mg] and [O/Mg] ratios of 2.4 and
1.4, respectively; the values of [C/Mg] and [O/Mg] for HE1327-2326 are
1.1 and 1.5.  That our models underreproduce the combined [CO/Mg]
values for these stars by at least 2 dex does not imply that they are
not representative of Population III SNe.  Since only [C/H] and [O/H]
ratios greater than -3.5 produce enough fine structure cooling to
trigger the rollover from Population III to II star formation
\citep[e.~g.][]{mbh03}, \citet{Frebel2007} have argued that any star
surviving until today with [Fe/H] $< -4$ \textit{must} have enhanced
abundances of C and/or O. Consequently, although perhaps much more
plentiful, Population III SNe that did not produce high [C/Fe] and
[O/Fe] ratios would not have imprinted their chemical signatures upon
a subsequent generation of stars that was long lived. More exotic
explosion mechanisms may not have been predominant at high redshift,
and may not even be necessary for reproducing the abundance patterns
in the more iron-poor stars yet found.  Fig \ref{fig:mpmp} shows that
for some stars, C, N, and O are mixed out to higher mass coordinate
than Mg and heavier elements.  It is possible that HE0107-5240 and
HE1327-2326 could both have formed from gas that was preferentially
enriched by these further-flung lighter elements.

\subsubsection{EMP Stars}
\begin{figure*}
\centering
\plotone{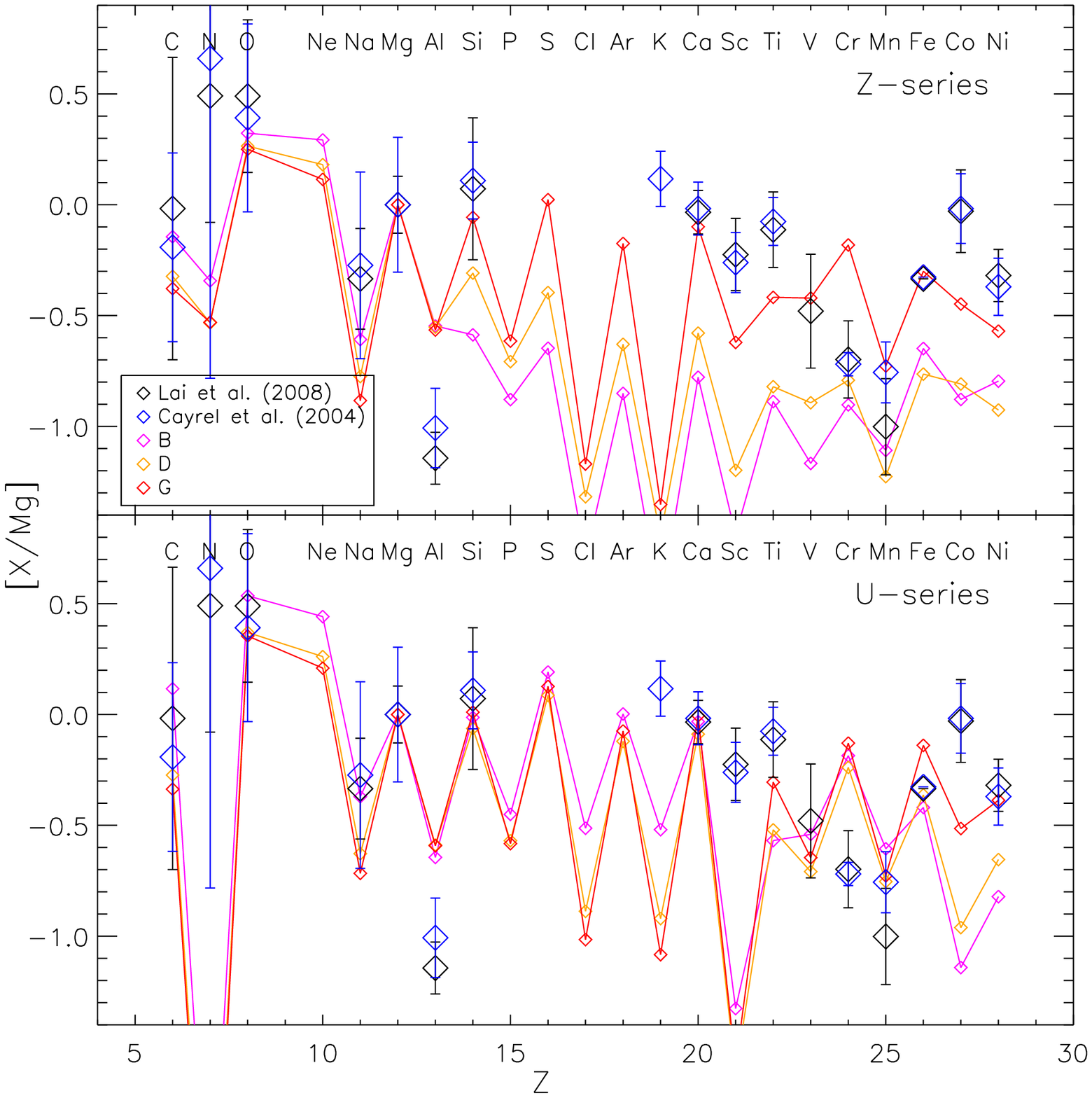}
\caption{IMF averages of yields compared to observations of EMP stars
  from \citet{Lai2008} and \citet{Cayrel2004}. IMF averaged yields are
  performed over a single explosion energy (B, D, or G) and over all
  masses in the sample. Higher-explosion energy rotating $Z=0$ stars
  reproduce EMP abundances well.  15 \Msun \ stars are needed to
  produce this good agreement with observations.}
\label{fig:empcomp}
\end{figure*}

Many more stars have been found with [Fe/H] values between -4 and -3
than below -4.5.  The abundances in these stars show less scatter than
in the small sample of three below -4.5.  Figure \ref{fig:hmpscat}
shows the average ratios of various abundances to Mg as a function of
[X/Mg] from the surveys of \citet{Lai2008} and \citet{Cayrel2004}.
Figure \ref{fig:empcomp} shows IMF averaged yields for 0.6, 1.2, and
2.4 Bethe explosions and $Z = 0$ and $10^{-4}$ \Zsun.  We used a Salpeter IMF,
\begin{equation}
 N(M)dM \propto M^{-\alpha}
\end{equation}
and the cannonical value for $\alpha$ of 2.35.

The fact that the u-series models don't produce enough nitrogen should
perhaps be ignored, since these models were constructed with yields
from non-rotating $Z = 0$ stars, which did not produce nearly the
nitrogen that was formed in the rotating $Z = 0$ \Zsun \ models.
Explosions of 1.2 and 2.4 Bethe (i.e. ``D'' and ``G'' models) for $Z =
0$ come closest to matching the abundances of the EMP stars.

An IMF-average over yields from the $Z = 0$ models provides a good fit
to the Cayrel and Lai data.  Aluminum is over produced by about 0.5
dex, and potassium is under produced by about 2 dex.  Scandium and
titanium are under produced by about 0.3 dex. Cobalt is under produced
by less than 0.5 dex.  But for all other elements our synthetic yields
fall within the error bars, or our yields from explosions of 1.2 or
2.4 Bethe bracket the data, demonstrating that SNe with energies
between 1.2 and 2.4 Bethe can account for the observed abundances.
However, 15 \Msun \ stars are needed in our IMF average to match the
abundance patterns in both the Lai and Cayrel data.  Spherical
explosions with energies above 2.4 Bethe are not required to match
abundance patterns in the EMP data sets.  While this does not rule out
the existence of hypernovae, it does suggest that they are not
required to explain the abundances in EMP stars.  Also, as we noted
earlier, our estimates of mixing are a lower bound. With more detailed
simulation physics like that in \citet{ Kifonidis2006}, it is likely
that more heavy elements would be produced at a given explosion energy
and that lower energy explosions alone would be sufficient to match
the EMP data.

Our IMF-averaged yields for rotating, zero-metallicity stars come
closer to the observed abundances in EMP stars than do the
one-dimensional parametrizations of \citet{Heger2008} or
\citet{Iwamoto2005}, the two-dimensional non-rotating spherical
explosions of \citet{Joggerst2009} or the jet models of \citet{
Tominaga2009}.  The angle-averaged yields of \citet{Tominaga2009} also
underestimate potassium by nearly 2 dex. Nitrogen, aluminum, scandium,
and cobalt are also under produced, to a far greater degree than in
our models.  Although the jet-induced explosion may better account for
the abundances in HE0107-5240 and HE1327-2326, an IMF average of
spherical explosion yields of rotating zero-metallicity stars is a
better fit to the \citet{Lai2008} and \citet{Cayrel2004} surveys.
These two HMP stars may well be singular, rare objects.  The more
comprehensive surveys of EMP stars likely reflect more typical
conditions and hence typical early supernovae. The good match of
IMF-weighted averages of supernovae yields to these EMP star data
suggests that supernovae of $\approx$ 15 \Msun \ with moderate
explosion energies of less than 2.4 Bethe were responsible for the
bulk of early enrichment.  Hypernovae may not be necessary to explain
these abundance patterns. Our results imply that moderately sized,
moderately energetic explosions produced most of the metals that went
on to form these EMP stars.

\subsection{Incorporation of SNe Ejecta into Metal-Poor Stars}

\begin{figure*}
\centering
\plotone{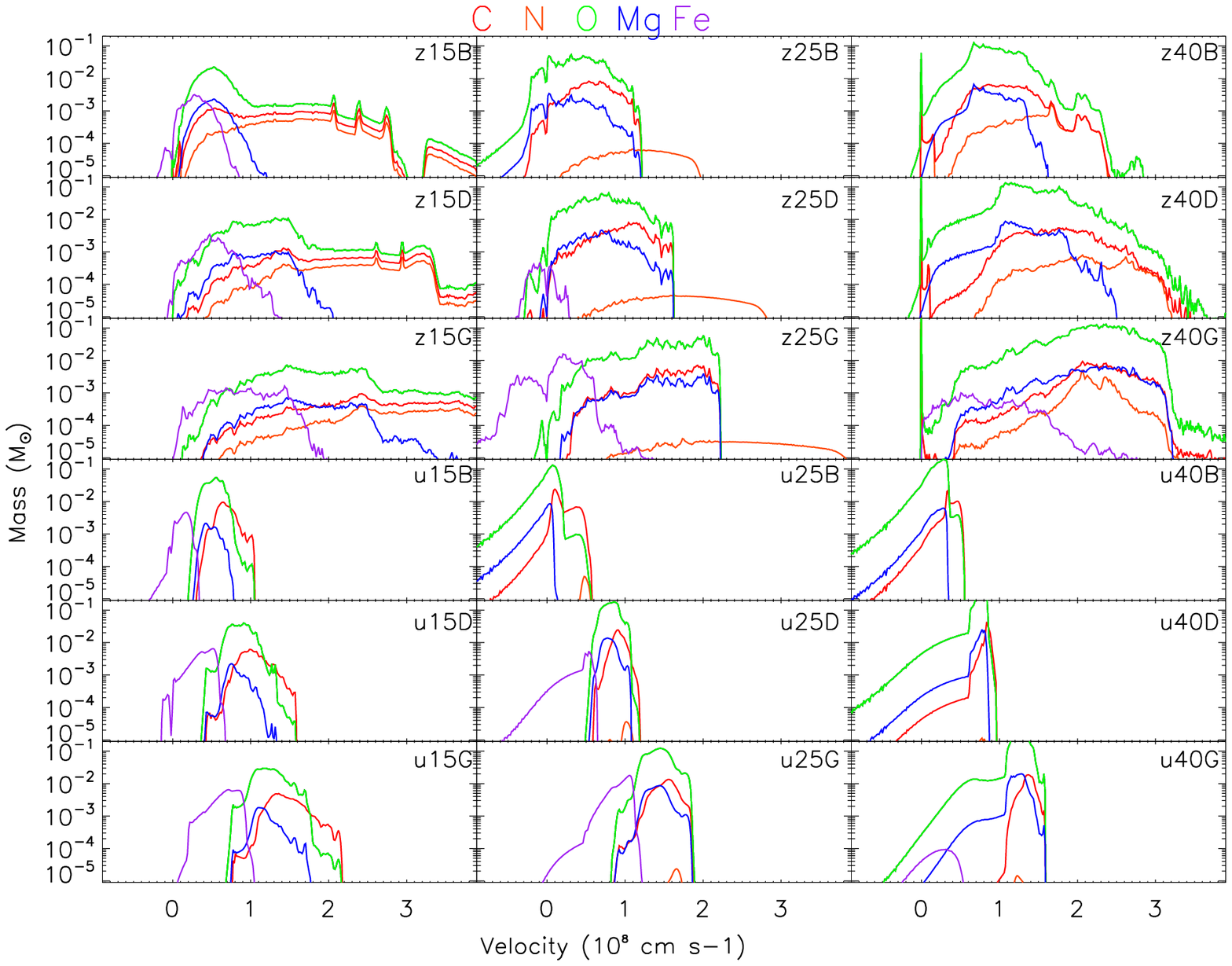}
\caption{Distribution of elements with velocity.  Progenitor mass 
  increases from left to right; explosion energy increases towards 
  the bottom of the figure; the first 3 rows correspond to $Z=0$
  stars, the bottom 3 rows correspond to $Z=10^{-4}$ stars.  The 
  more compact $Z=10^{-4}$ \Zsun \ stars demonstrate more fallback 
  (the long tails to the left of the peaks) than their $Z=0$ 
  counterparts.  Heavier elements are mixed our farther in velocity 
  space for $Z = 0$ than $Z=10^{-4}$.  Lower explosion energy leads 
  to more fallback in all models, especially as mass increases. In 
  all models in which Fe is ejected, it noticeably lags C, N, and O
  in velocity, implying that outer regions of the SNR will be 
  preferentially enriched with those elements at later times.}
\label{fig:mpvp}
\end{figure*}

Matching primordial nucleosynthetic yields directly to HMP and EMP
abundances is problematic because not all of the elements may have
been incorporated into subsequent metal-poor stars.  Intervening
hydrodynamical processes complicate the uptake of heavy elements into
second-generation stars on several disparate spatial scales.  First,
as seen in our models, fallback onto the central remnant deep within
the initial explosion can deprive later generations of certain
elements, particularly Fe.  On intermediate scales, HMP or EMP stars
may have formed promptly from gas clumps created in hydrodynamical
instabilities in primordial SN remnants (SNR) on 100 pc scales
\citep{wet08}.  If so, only the heavy elements that reach these
instabilities by the time of their formation are those that become
imprinted on the second stars.  Their presence throughout the SNR can
be traced back to their distribution in the early free expansion on
scales of 10$^{-3}$ - 10$^{-2}$ pc, which is well beyond those of our
models.  The velocity profile of the free expansion is nearly linear
with radius \citep[e.g.][]{tm99}, so gas near the center of the
explosion has much lower velocities than at greater radii.  If this
free expansion is not well mixed, heavier elements at small radii may
never reach the formation sites of dense clumps at later times and
appear in the resultant stars. Such differential mixing in the SNR can
skew the abundance patterns of the second generation and is yet to be
properly modeled.

However, a rough idea of how these elements will be disseminated at
later times can be obtained from their distribution with respect to
velocity in our models, which we show in Figure \ref{fig:mpvp}. The
velocities in this figure are those after all mixing has ceased, which
occurs at a different time for each star.  Again, it is apparent that
lower explosion energy SNe have greater fallback than do higher
explosion energy SNe; their iron cores are especially likely to fall
back.  For some stars, such as models u25B and z25B, u40B, u40D, z40B,
and z40D, iron does not appear on the plot because it has all accreted
onto the remnant.  In general, because they exhibit more mixing, the
z-series models show heavier elements mixed out to a higher percentage
of the velocity of the lighter elements than the u-series models.
Nitrogen, because it is distributed throughout the helium envelope of
the $Z = 0$ models, is mixed out to a higher velocity than the other
metals.  

Perhaps the most striking feature of these plots is that C, N, and O
all consistently reach similar velocities that are noticeably greater
than those of Fe.  The separation between these elements and Fe will
increase as the remnant expands homologously.  This trend suggests
that at later times regions of the ejecta at greater radii, which may
be the sites of prompt second star formation, will be preferentially
enriched with carbon, nitrogen, and oxygen.  Models z15B, z15D, and
z15G especially show C, N, and O separated in velocity from Mg and
Fe. C, N, and O are also separated from Mg and Fe in mass coordinate,
as can be seen in \ref{fig:mpmp}.  This distribution may increase
[CNO/Mg] while leaving [Fe/Mg] (as well as the ratios of element
heavier than Mg to Mg) relatively unchanged, bringing our yields more
into line with the abundances of HMP stars HE0107-5240 and
HE1327-2326. Future simulations of the explosion over larger spatial
scales and longer timescales are now being planned to evaluate these
processes.

\section{Conclusions}
\lSect{conclusions}

We have constructed two-dimensional simulations of the post-explosion
hydrodynamics of rotating Population III and II core-collapse SNe.  We
find that the main effect of rotation on the explosions is to alter
the structure of the stellar envelope through which they propagate.
Even a small amount of rotation was enough to expand the compact
stationary $Z = 0$ stars like those of \citet{Joggerst2009} from blue
to red.  On the other hand, rotation had less effect on the
presupernova structures of $Z=10^{-4}$ \Zsun \ stars, which remained
blue and displayed a degree of mixing comparable to the non-rotating
z-series models of \citet{Joggerst2009}.  Once rotation was introduced
in the progenitors, its actual magnitude had little bearing on the
final structure of the star, or hence on the ejecta and yield of the
SNe.  There were no significant differences between R=5 and R=10
models of equal mass and explosion energy.

The Z-series red supergiant models experience more mixing than the
blue U-series models, and the redness or blueness of a supernova
progenitor, not explosion energy, rotation rate, or mass, has the
largest effect on stellar yields.  We find that RT instabilities mix a
greater percentage of the star in the Z-series Population III models
because the reverse shock takes longer to travel through the larger
star, thus the prssure gradient is reversed for a longer period of
time and the RT instability has more time to develop.  The shorter
travel time for the reverse shock through the more compact u-series
models truncates the development of the RT instability in these stars
after a shorter time than their z-series counterparts, resulting in
less mixing.  Likewise, there is less fallback onto the central
remnant in the z-series explosions than in the blue u-series
Population II models, whose envelopes are more tightly bound.  At a
given mass and metallicity, higher explosion energies lead to more
mixing than lower energies, a trend that was most noticeable in the 40
\Msun \ runs.  More massive stars exhibited less mixing than
lower-mass stars.  This was most apparent in the blue u-series stars,
which experienced less mixing overall than their red z-series
counterparts.  Fallback increased with the mass of the progenitor and
fell with increasing explosion energy.  The 40 \Msun \ models with
explosion energies of 1.2 Bethe or less did not eject any iron, nor
did 25 \Msun \ models with 0.6 Bethe.

Our rotating progenitor explosion models can account for the abundance
patterns of the EMP stars and of one HMP star.  The abundances in
HE0557-4840, the least metal-poor of the HMP stars, with [Fe/H]=-4.75,
are modeled well by the yields of a 15 \Msun \ $Z = 0$ SN with 2.4
Bethe.  We are not able to replicate the high CNO to Fe ratios in the
more metal-poor HMP stars HE0107-5240 and HE1327-2326, although we do
reproduce their elements above Na with $Z = 0$ SNe models between 15
and 25 \Msun.  Enrichment by an AGB companion may explain the extra C,
N, O, and Na in HE 0107-5240 and HE1327-2326.  Although they explain
the abundance patterns in the two most iron-poor stars yet found,
jet-driven explosions may not have been the most prevalent type of
Population III supernova, just the one whose nucleosynthetic imprint
survives in HMP stars in the Galactic halo today.  The only stars with
[Fe/H] $< -5$ that could have formed with masses low enough to be
observable today were those with high [C/Fe] and [O/Fe]
\citep{Frebel2007, Frebel2009}.  Another explanation for the
high [CNO/Mg] ratios observed in these stars may lie in the final
distribution of elements in our simulations.  Models z15D and z25D,
provide the closest fits to elements heavier than Mg for HE0107-5240
and HE1327-2326, respectively.  These stars show C, N, and O separated
from Mg and heavier elements in both mass and velocity coordinates.
Preferential enrichment in these further-flung light elements may also
reproduce the chemical abundance patterns in these stars without the
need for a companion, which has not been observed despite continued
observations \citep{Frebel2008}.  Low-mass stars with [Fe/H] $<5$
might only have been able to form in a CNO-rich environment.

The assertion that stars like the ones in this paper represent the
dominant form of supernovae in the early Universe is bolstered by the
fact that our yields are a better match to the EMP stars than HMP
stars, which were likely enriched by a few generations of SNe whose
cumulative C, N, and O imprint was sufficient to trigger a rollover
from high-mass Population III stars to low-mass metal-poor Population
II stars.  Indeed, because the chemical makeup of EMP stars is the
product of a few generations (and therefore a better cross section) of
zero or very low metallicity explosions, the chemical abundances found
in EMP stars may be a better metric of those of $Z = 0$ and 10$^{-4}$
SNe in general.  An IMF average over spherical-explosion SN yields
from zero-metallicity rotating 15, 25, and 40 \Msun progenitors with
energies between 1.2 and 2.4 Bethe reproduces the abundance patterns
found by the \citet{Lai2008} and \citet{Cayrel2004} survey of EMP
stars with $-4 < $[Fe/H]$< -3$, with the exception of potassium, which
is too low by $\sim$ 2 dex in our models, and Al, which is too high by
$\sim$ 0.5 dex.  The fact that 15 \Msun \ progenitors must be included
in the IMF average suggests that the metal-free stars that contributed
the bulk of the metals to the early universe were of fairly low mass,
extending down to the lower limit predicted for Population III stars.
The abundance patterns of both HMP and EMP stars will soon come more
fully into focus as the \textit{Sloan Extension for Galactic
Understanding and Exploration-II} (\textit{SEGUE-II}) component of the
\textit{Sloan Digital Sky Survey-III} (\textit{SDSS-III}) uncovers
hundreds of thousands of low-mass metal-poor stars in the Galactic
halo.

While the uncertainty about the explosion mechanism and geometry of
core-collapse SNe means that these mixing estimates are likely lower
bounds, our essential conclusions are unchanged. Higher-mass stars may
experience more mixing, and less energetic blasts may eject amounts of
heavy metals similar to those in our models with more realistic
conditions for the first few seconds of the explosion. Hypernovae, or
explosions with 10 times the energy observed in the nearby Universe,
are not necessary to account for the abundances of the EMP stars:
moderately enhanced explosion energies of less than 2.4 Bethe suffice.
Our models are consistent with observations of abundances in EMP
stars, suggesting that jet-driven mechanisms may not be necessary to
explain these patterns.

Our conclusions assume that SN ejecta that escapes fallback onto the
remnant is uniformly mixed with the IGM and other SNR prior to
incorporation into EMP stars, which may not be the case.  Furthermore,
even though the velocity distributions of elements in our
zero-metallicity models are otherwise fairly uniform, Fe typically
lags C, N, and O.  Metals in our u-series models exhibit more
stratification in velocity space, suggesting that they will be
differentially imprinted onto subsequent generations of stars.
Simulations that follow the remnant out to greater spatial scales are
necessary to test this hypothesis and are now under development.
Mixing in three dimensions may also be different than in two, and will
also be investigated in future simulations.

\acknowledgments

Work at UCSC and LBL was supported in part by the SciDAC Program under
contract DE-FC02-06ER41438.  Work at LANL was carried out under the
auspices of the National Nuclear Security Administration of the
U.S. Department of Energy at Los Alamos National Laboratory under
Contract No. DE-AC52-06NA25396.  The simulations were performed on the
open cluster Coyote at Los Alamos National Laboratory. Additional
computing resources were provided on the Pleiades computer at UCSC
under NSF Major Research Instrumentation award number AST-0521566.

\bibliographystyle{apj.bst}
\bibliography{ms}

\end{document}